\documentclass[12pt,a4paper]{extarticle}
\usepackage[left=2cm,right=2.5cm,top=2.5cm,bottom=2.5cm]{geometry}
\usepackage{graphicx} 
\usepackage{float}
\usepackage{amsmath}
\usepackage{amssymb}
\usepackage{graphicx}
\usepackage{slashed}
\usepackage{xcolor}
\usepackage[colorlinks=true, allcolors=blue]{hyperref}
\usepackage{simpler-wick}

\usepackage{blkarray}
\usepackage{physics}
\usepackage{cite}

\DeclareFontShape{OT1}{cmr}{mx}{n}%
{<->cmr10}{}
\newcommand{\mytitlefont}{\fontseries{mx}\selectfont}
\DeclareMathAlphabet{\titlemath}{OT1}{cmr}{mx}{n}

\def\tilde{\widetilde}

\def\hat{\widehat}

\def\bar{\overline}

\def\d{\partial}

\def\1{{\mathds 1}}

\def\vol{\mathop{\rm vol}}
\DeclareMathAlphabet{\mathbfsf}{OT1}{cmss}{bx}{n}

\newcommand{\R}{{\mathbb R}}

\def\CD{{\mathcal D}}

\def\CK{{\mathcal K}}
\def\CL{{\mathcal L}}

\def\CN{{\mathcal N}}
\def\CO{{\mathcal O}}

\newcommand{\beq}{\begin{equation}\begin{aligned}}
\newcommand{\eeq}{\end{aligned}\end{equation}}

\newcommand{\bea}{\begin{eqnarray}}
\newcommand{\eea}{\end{eqnarray}}

\newcommand{\beqa}{\begin{eqnarray}}
\newcommand{\eeqa}{\end{eqnarray}}
\newcommand{\beqar}{\begin{eqnarray*}}
\newcommand{\eeqar}{\end{eqnarray*}}

\def\({\left(} \def\){\right)}
\def\[{\left[} \def\]{\right]}

\begin{document}


\begin{titlepage}

\begin{center}
			
~\\[0.9cm]
			
{\fontsize{26pt}{0pt} \mytitlefont Higher dimensional holography}
			
~\\[1cm]

Raquel Izquierdo Garc\'ia,$^{a}$$^{b}$\,\footnote{\href{mailto:rizquierdogarcia@perimeterinstitute.ca}{\tt rizquierdogarcia@perimeterinstitute.ca}}

~\\[0.5cm]

{\it $^{a}$Perimeter Institute for Theoretical Physics,\\ 
Waterloo, Ontario, N2L 2Y5, Canada}\

\bigskip

{\it $^{b}$Department of Physics and Astronomy, University of Waterloo,\\ 
Waterloo, Ontario, N2L 3G1, Canada}\
\end{center}

\vskip0.5cm
			
\noindent 

\makeatletter
\renewenvironment{abstract}{%
    \if@twocolumn
      \section*{\abstractname}%
    \else 
      \begin{center}%
        {\bfseries \normalsize\abstractname\vspace{\z@}}
      \end{center}%
      \quotation
    \fi}
    {\if@twocolumn\else\endquotation\fi}
\makeatother

  \begin{abstract}  
  \normalsize
The holographic dictionary is well developed for gravity in asymptotically anti de Sitter $A \left( AdS_{d+1} \right)$ spacetimes. However, this approach is limited, since many physically relevant configurations, such as \textit{bubbling geometries} dual to heavy operators, do not arise as an uplift of a lower-dimensional $A \left( AdS_{d+1} \right)$ solution. Instead, they are intrinsic solutions of ten- or eleven-dimensional supergravity. In this work, we address this problem by evaluating the on-shell Type IIB supergravity action, with a suitable boundary counterterm, on the bubbling solutions describing $1/2$ BPS surface operators in $\CN=4$ super Yang–Mills. This calculation exactly reproduces the surface operator Euler conformal anomaly, an observable known exactly in the field theory but previously inaccessible holographically. This example illustrates the need for a higher-dimensional approach to holographic renormalization applicable to general asymptotically $AdS_{d+1} \times  M_q$ geometries.
\end{abstract}

\vfill  

\end{titlepage}

\tableofcontents

\newpage

\section{Introduction}

Shortly after Maldacena's proposal of the duality between $\CN=4$ super Yang-Mills and String Theory in $AdS_5 \times S^5$ \cite{Maldacena:1997re}, the holographic dictionary for correlation functions was established \cite{Gubser:1998bc,Witten:1998qj}. This dictionary was developed for gravity theories in asymptotically anti de Sitter $A \left( AdS_{d+1} \right)$ spacetimes, where the boundary values of the bulk fields play the role of sources for the dual CFT operators. The generating functional of CFT correlators is obtained by evaluating the supergravity action on-shell with specific boundary conditions. Functional differentiation of the on-shell action with respect to the boundary sources then gives the CFT correlation functions of the corresponding local operators. This holographic principle has been successfully extended and tested in different contexts. Several generalizations to this set up involve 10d/11d supergravity in  $A(AdS_{d+1}) \times M^q$ where $M^q$ is some compact manifold \cite{Seiberg:1999xz,Gauntlett:2004yd,Gauntlett:2004zh,Klebanov:1998hh,Witten:1998zw}. In addition, the holographic dictionary has also been applied in the ``bottom-up'' approach to AdS/CFT where theories of gravity in $A(AdS_{d+1})$ spacetimes are related to a CFT living in the boundary without reference to a higher dimensional origin \cite{Hartnoll:2008kx,Takayanagi:2011zk,Kovtun:2004de,Lee:2008xf}.

The supergravity on-shell action suffers from IR divergences, which can be systematically removed  in asymptotically $AdS$ spacetimes using a method called \textit{holographic renormalization} \cite{Skenderis:2002wp,Henningson:1998ey,Henningson:1998gx,Kraus:1999di}. Even though the duality is believed to hold for the full theories defined in $AdS_{d+1} \times M^q$, the holographic dictionary has mostly been developed after Kaluza Klein reduction on the internal manifold and truncating the spectrum to a finite number of lower-dimensional fields. This approach usually requires that the $AdS_{d+1}$ supergravity theory is a consistent truncation of 10d/11d supergravity, ensuring that any solution of the lower-dimensional equations of motion uplifts to a solution of the higher-dimensional equations of motion \cite{duff:1984hn}.

However, many physically relevant solutions of 10d/11d supergravity do not arise as an uplift of a lower-dimensional $A (AdS_{d+1})$ solution. For example, when the geometry only  reaches $AdS_{d+1} \times S^q$ asymptotically, while having a topologically non-trivial interior \cite{Kraus:1998hv,Lin:2004nb,DHoker:2007mci,Gomis:2007fi,Yamaguchi:2006te,Gomis:2006cu,DHoker:2007hhe}. An important example are \textit{bubbling geometries}\footnote{The first example were the \textit{Bubbling AdS solutions} \cite{Lin:2004nb} dual to local operators with conformal dimension $\Delta = \CO (N^2)$. Many examples of topologically non-trivial solutions of  10d/11d supergravity describing the backreaction caused by the insertion of heavy  operators are known \cite{DHoker:2007mci,Gomis:2006cu,Gomis:2007fi,Lunin:2006xr,Yamaguchi:2006te,DHoker:2007zhm,DHoker:2007hhe,DHoker:2008rje,DHoker:2008lup }.}, which are dual to heavy operators. In general, asymptotically $AdS_{d+1} \times M^q$ geometries correspond to CFTs with defects or CFTs in a non-trivial state, and the standard tools of holographic renormalization often do not apply.

The \textit{Kaluza-Klein holography} program \cite{Skenderis:2006uy} developed by Skenderis and Taylor extends the usual tools of holographic renormalization to a more general framework. They propose a method to compute one-point functions probing any asymptotically $AdS_{d+1} \times S^q$ solution to supergravity. Their key insight is that, by considering a fluctuation around the $AdS_5 \times S^5$ vacuum and performing a non-linear Kaluza-Klein reduction in the asymptotic $S^5$, one can derive a 5d action that includes the full tower of massive modes, organized order by order in the number of the fields. The expectation values can then be computed from this five-dimensional $AdS_5$ action using the usual holographic dictionary \cite{Witten:1998qj,Gubser:1998bc} and the standard techniques of holographic renormalization \cite{Skenderis:2002wp,Henningson:1998ey,Henningson:1998gx,Kraus:1999di}.



The 5d action derived in \textit{Kaluza-Klein holography} depends, in general, on an infinite number of fields. As a consequence, no systematic procedure is currently available to compute the on-shell action within this framework. Nevertheless, correlation functions of the form $\langle \CO_{\text{heavy}} \CO_{\text{light}} \rangle$ remain accessible, since they are obtained from functional derivatives of $S^{5d}_{on-shell}$. For an operator $\CO_{\text{light}}(x)$ of conformal dimension $\Delta$, only a finite number of terms in the 5d action contribute to $\langle \CO_{\text{heavy}} \CO_{\text{light}} \rangle$, despite the non-linear KK map\footnote{See \cite{Skenderis:2006uy} or section 2 of \cite{Skenderis:2007yb}. In section \ref{sectioncutoff}, when we compare our cut-off prescription with theirs we also summarize this procedure. }. 
This framework has been successfully applied to several \textit{bubbling geometries}, such as three point functions and four-point functions of $1/2$ BPS local operators \cite{Skenderis:2007yb, turton:2024afd,aprile:2024lwy}, correlation function of $1/2$ BPS Wilson loop and $1/2$ BPS surface operator with local operators \cite{gomis:2008qa,Drukker:2008wr}, among others. However, observables sensitive to the finite part of the on-shell action probe the full non-linear Kaluza-Klein map. To our knowledge, there is no prescription for computing such quantities within a 5d truncation involving finitely many fields.


Alternatively, one can try to give a prescription on how to evaluate the 10d/11d supergravity action and provide a higher dimensional approach to holography. Some early works exploring this are \cite{Taylor:2001fe,Taylor:2001pp}, but a systematic framework for holography applicable to general asymptotically $AdS \times X$ spacetimes remains an open problem. In this work, we adopt this higher dimensional perspective and directly compute the on-shell 10d supergravity action.


Superconformal non-local operators preserving a large amount of supersymmetry are the perfect setting to address this problem, since certain observables that have been computed at strong coupling can be used to test the holographic duality in cases where a fully backreacted \textit{bubbling geometry} is known. In particular,  some observables correspond directly to the finite part of the regularized on-shell action and cannot be computed using Kaluza-Klein holography. In this work, we focus on the Euler anomaly $b$ of the $1/2$ BPS surface operator $\CO_\Sigma$ of $\CN=4$ super Yang-Mills, an observable that has been computed exactly in the field theory but lacking a holographic calculation\footnote{The holographic description of $\CO_\Sigma$ when $b= \CO(N) $ is given in terms of probe D3 branes on $AdS_5 \times S^5$\cite{Drukker:2008wr}. The on-shell action of a single probe D3 brane, corresponding to $b=6N-6$, has been computed and showed to match the leading large $N$ result in field theory \cite{Gutperle:2020gez,Jiang:2024wzs}. Subleading terms in $1/N$ are being investigated from the probe brane perspective \cite{Jiang:2024wzs}.}.


In the regime where $b = \CO(N^2)$, the holographic dual of $\CO_\Sigma$ is a \textit{bubbling geometry} \cite{Gomis:2007fi}, and $b$ must be the coefficient of a logarithmic singularity in the type IIB supergravity on-shell action. In this work, we evaluate the type IIB supergravity action in this background and reproduce the field theory value of $b$ by adding a suitable 9d counterterm. 

Several fully backreacted geometries dual to supersymmetric defects are known, many of which exhibit an analogue of the Euler anomaly of the surface operator. Whether the holographic prescription that we propose here extends to these cases is an interesting question, which we leave for future investigation.

The rest of the paper is organized as follows. In section \ref{section:surfaceop}, we review the definition of 1/2 BPS surface operators in $\CN=4$ SYM and their conformal anomalies. In section \ref{sec:dualholography}, we review the backreacted geometries dual to $\CO_\Sigma$ and how to define a Fefferman-Graham cut-off. We also discuss the cut-off choice and compare it with the one used in \cite{Skenderis:2007yb,Drukker:2008wr}. In section \ref{sec:onshellactionsec}, we summarize the calculation of the type IIB on-shell action from \cite{Gentle:2015ruo}. Section \ref{sec:counterterm} proposes a cosmological constant counterterm in the 9d regulated boundary. In \ref{conclusions} we conclude and discuss some future work. Some lengthy expressions technical details are collected in the appendices.

\section{\texorpdfstring{1/2-BPS surface operators in $\mathcal{N}\hspace{-0.1em}=$ 4 super Yang-Mills}{1/2-BPS surface operators in N=4 super Yang-Mills}}
\label{section:surfaceop}

1/2-BPS surface operators of $\CN=4$ super Yang–Mills were constructed by Gukov and Witten \cite{Gukov:2006jk}. Preserving half of the supersymmetries of $\CN=4$ SYM constrains the support $\Sigma$ of these operators to be either a plane or a sphere. These conformal surface operators break the superconformal symmetry of $\mathcal{N}=4$ SYM into $PSU(1,1|2) \times PSU(1,1|2) \times U(1) \subset PSU(2,2|4)$. The corresponding bosonic Euclidean subgroup is $SO(1,3) \times SO(2) \times SO(4)$, where $SO(1,3)$ is the 2d global conformal group acting on $\Sigma$, $SO(4) \subset SO(6)$ is an $R$ symmetry subgroup, and $SO(2)_D = \text{diag} (SO(2)_R \times SO(2)_\perp)$ is a diagonal combination of rotation in the transverse plane to $\Sigma$ and $R$-symmetry rotation\footnote{Usually, in the definition of conformal defects it is asked that the transverse rotations to the defect are preserved. However, this is not the case here unless $\beta = \gamma = 0$, ass discussed below.}.

Surface operators deform the theory by introducing a charged probe string, whose worldsheet corresponds to their support $\Sigma$. The ones we consider here are ``disorder'' operators, since they cannot be constructed from the fields in the path integral of the theory\footnote{This description characterizes the surface operator as singularities of the fields near $\Sigma$. Alternatively, surface operators can be defined by coupling a 2d theory supported on $\Sigma$ to the bulk 4d theory \cite{Gukov:2008sn}. For a review, see \cite{Gukov:2014gja}.}. Instead, they are defined by prescribing a singular boundary conditions for some fields near $\Sigma$. The singularities are labeled by a Levi subgroup $\mathbb L= \prod_{l=1}^M U(N_l) \subset U(N)$ in $\Sigma$, and $4M$ real parameters $(\alpha_l,\beta_l,\gamma_l,\eta_l)$\footnote{Appendix A of \cite{Gomis:2007fi} shows which are the supersymmetries of $\CN=4$ SYM are preserved by $\CO_{\mathbb{R}^2}$.}. For readability, we will omit most labels and denote the surface operator simply as $\CO_\Sigma$.

The gauge field has a non-abelian vortex singularity near $\Sigma$, characterized by a prescribed holonomy around the defect: 
\begin{equation}
\label{backgroundA}
A= \alpha \, d \psi=
    \left(\begin{array}{cccc}
         \alpha_1  \mathbb{I}_{N_1}&0 &s & 0  \\
         0& \alpha_2  \mathbb{I}_{N_2}&s & 0 \\
         \vdots &\vdots &\ddots&\vdots \\
         0&0&s&\alpha_M \mathbb{I}_{N_M}
    \end{array} \right) d \psi, 
\end{equation}
where $\psi$ is the polar angle in the transverse space to $\Sigma$\footnote{Alternatively, if $z= r e^{i \psi}$ is a complex coordinate in the plane transverse to $\Sigma$, $A_z = \frac{\alpha}{2 i z}$, which is compatible with scale invariance. The parameter $\alpha$ can be interpreted as the Aharonov-Bohm phase for a charged particle that goes around the defect.}, the parameters $\alpha_l$ take values in the unit circle and $\mathbb{I}_N$ is the $N \times N$ identity matrix, so that $\alpha$ is in the maximal torus of $U(N)$. One of the complex scalars of $\mathcal{N}=4$ super Yang–Mills also develops a singularity compatible with scale invariance given by:
\begin{equation}
\label{backgroundphi}
\Phi=\frac{1}{\sqrt{2}\,z}
    \left(\begin{array}{cccc}
         (\beta_{1}+i\gamma_1) \mathbb{I}_{N_1}&0 &s & 0  \\
         0& (\beta_{2}+i\gamma_2) \mathbb{I}_{N_2}&s & 0 \\
         \vdots &\vdots &\ddots&\vdots \\
         0&0&s&(\beta_{M}+i\gamma_M) \mathbb{I}_{N_M}
    \end{array} \right),
\end{equation}
where $\beta_l, \gamma_l$ are real numbers. Two-dimensional $\theta$-angles can be introduced on the surface $\Sigma$ for each $U(1)$ factor in a way consistent with the symmetries. They can be combined into an $\mathbb L$-invariant matrix:
\begin{equation}
\label{thetas}
\eta=
    \left(\begin{array}{cccc}
         \eta_1 \; \mathbb{I}_{N_1}&0 &s & 0  \\
         0& \eta_2  \mathbb{I}_{N_2}&s & 0 \\
         \vdots &\vdots &\ddots&\vdots \\
         0&0&s&\eta_M \mathbb{I}_{N_M}
    \end{array} \right).
\end{equation}
The parameters $\eta_I\simeq \eta_I+1 $ are also circle valued, taking values on the maximal torus of the
$S$-dual or Langlands dual gauge group  $^LG$, where  $^LG=G=U(N)$. The matrices $\alpha,\beta,\gamma,\eta$ are invariant under $\mathbb{L}$.

Weyl anomalies of conformal surface operators are much richer than those of 2 dimensional CFTs \cite{Graham:1999pm,Henningson:1999xi,Schwimmer:2008yh}. Under a Weyl transformation $\delta \sigma$ of a background $d$-dimensional metric $g_{\mu\nu}$, the expectation value of a surface operator transforms like \cite{Graham:1999pm,Henningson:1999xi,Schwimmer:2008yh}  
\beq 
\delta_\sigma\log \langle {\cal O}_\Sigma \rangle= \frac{1}{24 \pi}\int_\Sigma d^2x \;\sqrt{h}\;\delta\sigma\;\left(b \, R_{\Sigma}+c_1 \, g_{mn}  h^{\mu\sigma}h^{\nu\rho} \hat K^m_{\mu\nu} \hat K^n_{\rho\sigma}-c_2 \, W_{\mu\nu\rho\sigma}h^{\mu\rho}h^{\nu\sigma}\right)\,,
\label{Weylanom}
\eeq
where $h_{\mu\nu}$ is the induced metric on $\Sigma$, $R_\Sigma$ is the Ricci scalar of $\Sigma$, and $\hat K^m_{\mu\nu}$ and $W_{\mu\nu\rho\sigma}$ are the traceless part of extrinsic curvature and pullback of Weyl tensor respectively. $b$ is a type A anomaly while $c_1,c_2$ are type B according to the classification of \cite{Deser:1993yx}.

In analogy with the conformal dimension of local operators, the scaling weight $h$ of different defects can be defined through their correlation function with the stress-tensor. It can be written in terms of $c_2$ \cite{Bianchi:2015liz}. Alternatively, $c_1$ is related to the two-point function of the displacement operator \cite{Bianchi:2015liz}. The anomaly $b$ multiplying the Euler characteristic of $\Sigma$ is known to be monotonic under defect RG flows  \cite{Jensen:2015swa, Wang:2020xkc}.

In this work, we focus on computing the anomaly coefficient $b$ of $\CO_{S^2}$ holographically. If we take \eqref{Weylanom} for the case $\Sigma = S^2$ and flat ambient spacetime, we can integrate it and see how $b$ is directly related to the expectation value of the surface operator:
\beq 
\langle \CO_{S^2} \rangle  = \left(\frac{r}{r_0}\right)^{\frac{b}{3}},
\eeq
where $r$ is the radius of $\Sigma = S^2$ and $r_0$ is a scheme dependent scale\footnote{One can study $\langle \CO_\Sigma\rangle$ in $\mathbb R^4$ or $S^4$, they will differ by a contribution of the bulk $a$ conformal anomaly of $\CN=4$ SYM, this is well understood, see \cite{Chalabi:2020iie,Choi:2024ktc}.}. The Euler anomaly $b$ is scheme independent and,analogous to the central charge in 2d CFTs, is believed to count defect degrees of freedom. 


It has been proven that the $b$ anomaly of the $1/2$ BPS surface operator of $\CN=4$ SYM with Levi group $\mathbb L$ is \cite{Jensen:2018rxu,Chalabi:2020iie,Wang:2020xkc,Choi:2024ktc}
\beq 
\label{eq:banomaly}
b = 3 \left( \dim G - \dim \mathbb L \right).
\eeq

Then, 
\beq 
\log \langle \CO_{S^2} \rangle = \left( N^2 - \sum_{l=1}^M N_l^2 \right) \; \log \left( r \Lambda \right).
\eeq
Via the holographic dictionary, our goal is to reproduce this result from gravity by showing that
\beq 
S_{supergravity}^{on-shell}=\left( N^2 - \sum_{l=1}^M N_l^2 \right) \; \log \left( r \Lambda \right),
\eeq
where $S_{supergravity}^{on-shell}$ is the on-shell type IIB supergravity action evaluated on the geometry dual to the surface operator $\CO_{S^2}$.

\section{Holographic dual of \texorpdfstring{$\CO_\Sigma$}{COΣ}}
\label{sec:dualholography}


As noted by Gukov and Witten \cite{Gukov:2006jk}, the brane system studied in \cite{Constable:2002xt}, involving a stack of $N$ D3 branes intersecting a stack
of $N'$ D3 branes along a two-dimensional surface $\Sigma$, provides a string theory realization of $ \CO_\Sigma $. When $\mathbb L = U(N-1) \times U(1)$, the appropriate holographic description is given by a single probe D3 brane with $AdS_3 \times S^1$ worldvolume, which intersects the boundary of $AdS_5 \times S^5$ along $\Sigma = S^2$ or $\R^2$ \cite{Drukker:2008wr}. Fully backreacted geometries dual to these surface operators were constructed in \cite{Gomis:2007fi}. These bubbling geometries are smooth, horizonless, and asymptotically $AdS_5 \times S^5$. They describe the fully backreacted geometry of the intersecting D3 branes in the regime $b = \mathcal{O}(N^2)$, corresponding to $N' = \mathcal{O}(N)$ in the brane setup of \cite{Constable:2002xt}, where backreaction can no longer be neglected \cite{Drukker:2008wr,Gomis:2007fi}. The data characterizing $\CO_\Sigma$ -- the Levi subgroup $\mathbb L$ and the parameters $(\alpha_l,\beta_l,\gamma_l,\eta_l)$ -- completely determine the corresponding bubbling geometry. 

These geometries are obtained by imposing the $PSU(1,1|2) \times PSU(1,1|2) \times U(1)$ symmetry of $\CO_\Sigma$ in type IIB supergravity. In particular, the bosonic symmetry $SO(2,2) \times SO(4) \times SO(2) $ constrains the $10d$ metric to take the form of an $AdS_3 \times S^3 \times S^1$ fibration over a three-dimensional manifold $X$. Similar constraints apply to the five-form flux. These are the only two fields turned on in these solutions\footnote{In addition to the constant axio-dilaton that is mapped to the complexified gauge coupling.}:
\beq 
\label{eq:bubblinggeometry}
ds^2 &= f_1^2 ds^2_{AdS_3} +f_2^2 ds^2_{S^3} +f_3^2 \left(d\psi + V \right)^2 + f_3^2 ds^2_{X}, \\
F_{(5)} &= F \wedge \text{vol}_{AdS_3} + \tilde F \wedge  \text{vol}_{S^3}.
\eeq

Bubbling geometries dual to supersymmetric operators are typically constructed by solving the Killing Spinor equations and Bianchi identities for the field strengths, rather than the supergravity equations of motion. In $\CN=4$ SYM, the bubbling geometries dual to surface operators are constructed by double analytic continuation of the geometries describing the two point function of $1/2$ BPS local operators \cite{Lin:2004nb}, which preserve $SO(4) \times SO(4) \times R$. 


That analysis showed that the warp factors $f_I, \; I=1,2,3$ and $F, \tilde F$ have a particularly simple form and can be written in terms of $z$, a function on $X$, which obeys a differential equation. The 10d metric is given by
\beq 
\label{eq:metricbubblingsurface}
ds^2&=y \sqrt{\frac{2z+1}{2z-1}} \; ds^2_{AdS_3} + y \sqrt{\frac{2z-1}{2z+1}} \; ds^2_{S^3} +\\
&+\frac{2y}{\sqrt{4z^2-1}} \, (d \chi + V)^2 + \frac{\sqrt{4z^2 -1}}{2y} ds^2_X, 
\eeq
where 
\beq 
ds^2_X =dy^2 + dx_i \, dx_i
\eeq
is the metric on $X = (\mathbb{R}^3)^+$ and $V$ is a 1-form on X satisfying\footnote{Note that this equation defines $V$ up to an exact form.}
\beq 
\label{eq:difeqforV}
d \, V = \frac{1}{y} \star_X dz.
\eeq
 
The 3d manifold $X$ is non-compact and has a boundary located at $y=0$. The coordinate $y$ is the product of the radius of $AdS_3$ and $S^3$ \eqref{eq:metricbubblingsurface}. Regularity of the geometry requires the $S^3$ to shrink continuously to a point at $y=0$, which imposes the boundary condition $z(0,\vec x)= 1/2$ on the differential equation \eqref{eq:difeqforV}. Nontrivial solutions are obtained by introducing sources\footnote{Note that the LHS can be written like $y \; d\,(d\, V) $, which means that $dV$ is not closed at the location of the sources.}: 
\beq 
\label{eq:difeqnchargedistribution}
    \d_i \d_i \, z(y,\vec x) + y \, \d_y \left( \frac{1}{y}\,  \d_y \, z(y,\vec x) \right) = \sum_{l=1}^M Q_l \; \delta(y-y_l) \, \delta^2(\vec x - \vec x_l) . 
\eeq

The function $z$ and the one-form $V$ can be determined explicitly in terms of a configuration of $M$ ``charged particles'' in $X$, each  with positions $(y_l,\vec x_l)$ and charges $Q_l$. The general solution takes the form
\beq
\label{eq:zandV}
&z(y,\vec x)= \frac{1}{2} + \sum_{l=1}^M z_l(y,\vec x), \quad z_l(y,\vec x) = \frac{Q_l}{4 \pi y_l} \left( \frac{(\vec x- \vec x_l)^2 + y^2+y_l^2}{\sqrt{\left((\vec x- \vec x_l)^2 + y^2+y_l^2\right)^2 - 4 y^2 y_l^2}} - 1 \right) ,\\
&V_I dx^I = - \sum_{l=1}^M \sum_{I,J=1}^2 \frac{Q_l}{4 \pi y_l} \; \epsilon_{IJ} \;\frac{(x_J - (x_{l})_J)}{ (\vec x - \vec x_l)^2 } \frac{(\vec x- \vec x_l)^2 + y^2-y_l^2}{\sqrt{\left((\vec x- \vec x_l)^2 + y^2+y_l^2\right)^2 - 4 y^2 y_l^2}} dx^I.
\eeq
Since $z$ diverges at the location of the charges, the  $\chi$ circle in the metric \eqref{eq:metricbubblingsurface} shrinks at these points. Requiring the absence of a conical singularity around $(y,\vec x) = (y_l, \vec x_l )$ fixes the charges to be $Q_l = 2\pi y_l$. 
 
After imposing the Killing Spinor equation and the Bianchi identity for $F_{(5)}$, it can be shown that $F_{(5)}$ only depends on $z$ and $V$: 
\beq 
\label{eq:F_5bubblingsurface}
F_{(5)} &=- \frac{1}{4} \left[ d \left( y^2 \frac{2z + 1}{2 z -1} (d\chi +V) \right) - y^3 \star_X d \left( \frac{1}{y^2} \left(z+\frac{1}{2} \right) \right) \right] \wedge {\vol}_{AdS_3}+ \\
&-\frac{1}{4} \left[ d \left( y^2 \frac{2z - 1}{2 z +1} (d\chi +V) \right) - y^3 \star_X d \left( \frac{1}{y^2} \left(z-\frac{1}{2} \right) \right) \right] \wedge {\vol}_{S^3}.
\eeq


The geometry \eqref{eq:metricbubblingsurface} is topologically non-trivial. In particular, there are $M$ topological five-spheres supported by flux. These $S^5_l$ are built as a $S^3 \times S^1$ fibration over the interval joining $(0,\vec x_l)$ --where the $S^3$ smoothly shrinks to a point-- and $(y_l, \vec x_l)$ --where the $S^1$ does\footnote{Imposing that there is no conical singularity around $y=y_l$ fixes $Q_l = 2\pi y_l$. Note that this makes the differential equation \eqref{eq:difeqnchargedistribution} Laplace's equation in Poincar\'e coordinates of euclidean $AdS_3$ with point-like sources.}--. Imposing flux quantization gives a relation between the partition of $N$ defining the Levi subgroup $\mathbb L$ of the surface operator and the position of the particles \cite{Gomis:2007fi}:
\beq 
\label{eq:Nlintermsofyl}
N_l = \frac{y_l^2}{4 \pi l_P^4} = \frac{y_l^2}{L^4}N.
\eeq

This quantizes the $y_l$, which also must satisfy the constraint\footnote{One can see that this constraint corresponds to the metric \eqref{eq:metricbubblingsurface} being asymptotically $AdS_5 \times S^5$ with radius $L$.}
\beq 
\sum_{l=1}^M y_l^2  = L^4 .
\eeq

The location of the $M$ ``particles'' and the continuous parameters $(\beta,\gamma)$ labeling $\CO_\Sigma$ are mapped in the following way \cite{Gomis:2007fi}\footnote{Note that $\vec x_l$ corresponds to the location of the $N'$ D3 branes in the transverse space (which is filled by the $N$ D3 branes). This can be shown in the probe limit $\mathbb L = U(N-1) \times U(1)$ \cite{Drukker:2008wr}.}: 
\beq
(\beta_l,\gamma_l)=\frac{\vec x_l}{2\pi l_s^2}.
\eeq

Additionally, the non-contractible circle in the boundary ($\chi \sim \chi + 2\pi$), which continuously caps-off at the location of each source, corresponds to a non-trivial disk $D_l$. This means that the configuration in \eqref{eq:bubblinggeometry} is not unique, as one also needs to specify the holonomy of the two-form potentials of type IIB supergravity around the $M$ non-trivial disks. These holonomies are mapped in the holographic dictionary to the continuous parameters $\{\alpha_l, \eta_l\}$\cite{Gomis:2007fi}:
\beq
 \alpha_l =\int_{D_l}B_{\text{NS}}, \qquad \qquad \eta_l= \int_{D_l}B_{\text{R}}.
\eeq

This solution is asymptotically $AdS_5 \times S^5$, even though it is not manifest in the coordinates \eqref{eq:metricbubblingsurface}. In order to see this, it is useful to study the vacuum solution. The metric \eqref{eq:metricbubblingsurface} reduces to $AdS_5 \times S^5$ when $M=1$: there is only one particle located at $(y_0,\vec x_0)$. Using the change of coordinates\footnote{We can see in \eqref{eq:metricbubblingsurface} that $y, \vec x$ have dimensions of $\text{length}^2$. Note that the condition \eqref{eq:Nlintermsofyl} implies $y_0=L^2$.} \cite{Lin:2005nh}
\beq
\label{changeofcoords}
\chi &= \frac{1}{2}( \psi - \phi), \quad \alpha =\psi + \phi,\\
y &= y_0 \, \sqrt{\rho^2+1} \, \cos \theta, \\
x_1 &= (x_1)_0+ y_0 \, \rho \, \sin \theta \cos \alpha, \\
x_2 &= (x_2)_0+ y_0 \, \rho \, \sin \theta \sin \alpha, \\
\eeq
where $\rho \in (0, \infty), \; \theta \in \left[0, \pi/2\right] , \; \psi \in \left[0, 2\pi\right) $ and $\; \psi \in \left[0, 2\pi\right)$. Then, the function $z$ and the 1-form $V$ take a  simple form 
\beq 
z^{(0)}&= \frac{\cos ^2\theta +\rho ^2+1}{2 \left(\sin ^2\theta +\rho ^2\right)}, \\
V^{(0)}&= \frac{ \left(\rho ^2-\sin ^2\theta \right)}{2 \left(\sin ^2\theta +\rho ^2\right)}(d\psi +d\phi ).
\eeq
The metric \eqref{eq:metricbubblingsurface} is $AdS_5 \times S^5$ where $AdS_5$ has an $AdS_3 \times S^1$ slicing and $S^5$ is written as an $S^3 \times S^1$ fibration over an interval, and $F_{(5)}$ is proportional to the volume form of $AdS_5$ and $S^5$: 
\beq  
\label{eq:metricbubblingvacuum}
ds^2 =& L^2 \left( \frac{d\rho^2}{\rho^2 + 1} + (\rho^2 + 1) ds^2_{AdS_3} + \rho^2 d\psi^2 + d\theta^2 + \sin^2 \theta \, d\phi^2 + \cos^2 \theta ds^2_{S^3} \right), \\
F_{(5)}=& - L^4 \left( \rho  \left(\rho ^2+1\right) d\rho \wedge d\psi \wedge {\vol}_{AdS_3} + \sin \theta \cos ^3\theta d\theta \wedge d\phi \wedge {\vol}_{S^3}\right)
\eeq
where we have identified $y_0=L^2$. 

The conformal boundary of this spacetime is $AdS_3 \times S^1$. This means we are studying\footnote{Indeed, there are no bubbling solutions known in which the conformal boundary is $\mathbb{R}^4$.} $\CN = 4$ super Yang-Mills on $AdS_3 \times S^1$, not in $\mathbb R^4$ or $S^4$. However, $AdS_3 \times S^1$ and $\mathbb R^4$ are conformally related, and the former is the natural geometry for studying conformal surface operators in 4 dimensions. The surface operator $\CO_\Sigma$ is at the boundary of $AdS_3$, and the choice of coordinates determines the geometry of $\Sigma$. Poincar\'e coordinates 
\beq 
ds^2_{AdS_3} = {1 \over z^2} \left( dz^2 + dl^2 + dm^2 \right)
\eeq
give $\Sigma = \R^2$ when $z \to 0$ while global coordinates\footnote{As we have seen before, $\log \langle \CO_{S^2} \rangle$ has a logarithmic singularity depending on the radius of $S^2$, this will correspond in gravity to a logarithmic singularity in the on-shell action proportional to the volume of $AdS_3$.}
\beq 
ds^2_{AdS_3} = d r^2 + \sinh^2 r d\Omega_2^2
\eeq
give $\Sigma = S^2$ when $r \to \infty$.

It is clear that we can reach the boundary of \eqref{eq:metricbubblingvacuum} in two different ways. Away from the defect, by taking $\rho \to \infty$ with $r=r_0$ or $z=z_0$ fixed, or at a point on the defect, by taking $r \to \infty$ or $z \to 0$ while keeping $\rho$ fixed\footnote{See figure 1 in \cite{Jensen:2013lxa}, $z^{\text{here}}=Z^{\text{there}},\; \rho^{\text{here}} \sim x^{\text{there}}$}. This is always the case for holographic duals of non-local operators or boundaries and it makes it difficult to define a Fefferman-Graham expansion \cite{fefferman1985conformal}, which is needed to compute correlation functions in the gravity side of the AdS/CFT correspondence. We expand on this problem in the next subsection.

\subsection{Asymptotic expansion and Fefferman-Graham cut-off}
\label{sectioncutoff}

Even though defects in CFTs and their holographic duals have been studied extensively over the past twenty years, there is still not a good understanding of  the correct regularization cut-off or Fefferman-Graham expansion in the presence of defects. The holographic dual of a $p$-dimensional conformal defect in a $d$-dimensional CFT involves an $AdS_{d+1}$ with $AdS_{p+1}$ slicing. As mentioned above, this implies that there are two different ways to reach the boundary. In this context, we want to reach the boundary away from the defect. However, it is not well understood when does the Fefferman-Graham prescription we propose below break down or what is the physical meaning of such prescription. 

This problem was first encountered in BCFT, where a prescription for Janus solutions fixed a Fefferman-Graham expansion preserving symmetries left unbroken by the conformal boundary \cite{Papadimitriou:2004rz,Estes:2012nx}. This approach was later generalized to any $AdS^a \times S^b$ fibration of $AdS_{d+1}$ dual to non-local operators of any dimension \cite{Jensen:2013lxa}. It has also been used in several bubbling geometries to compute holographic entanglement entropy of spheres intersecting defects by computing minimal areas, using the Ryu-Takayanagi proposal \cite{Ryu:2006bv}. This has been done for Wilson loops \cite{Gentle:2014lva} and surface operators \cite{Gentle:2015ruo} in $\mathcal{N}=4$ SYM, and Wilson surfaces in the 6d $\CN = (2,0)$ theory, among many other examples. 

In this section, we review the choice of cut-off and asymptotic expansion used in \cite{Gentle:2015ruo} and compare it with the approach taken in \cite{Skenderis:2007yb,Drukker:2008wr}.

The main idea is to write the general bubbling solution \eqref{eq:metricbubblingsurface} as a perturbation of the vacuum $AdS_5 \times S^5$ solution \eqref{eq:metricbubblingvacuum}. The most useful change of coordinates exploits the overall translation symmetry in the $\vec x$ plane, by expanding around the center of mass \cite{Drukker:2008wr}
\beq
{\vec x \, }^{(0)} = \sum_{l=1}^M y^2_l \, \vec x_l , \quad (y^{(0)})^2= \sum_{l=1}^M y^2_l = L^4.
\eeq

The change of coordinates is given by
\beq
\label{changeofcoordsgeneral}
\chi &= \frac{1}{2}( \psi - \phi), \quad \alpha =\psi + \phi,\\
y &= L^2 \, \sqrt{\rho^2+1} \, \cos \theta, \\
x_1 &= (\vec x^{(0)})_1+ L^2\, \rho \, \sin \theta \cos \alpha, \\
x_2 &=  (\vec x^{(0)})_2+ L^2\, \rho \, \sin \theta \sin \alpha, \\
\eeq
where $\rho \in (0, \infty), \; \theta \in [0, \pi/2] , \; \psi \in [0, 2\pi) $ and $\; \psi \in \[0, 2\pi\)$.

It is useful to define the momenta \cite{Gentle:2015ruo} 
\beq 
\label{eq:momenta}
m_{abc} = \sum_{l=1}^M y_l^a x_{l1}^b x_{l2}^c,
\eeq
which will appear in the asymptotic expansions of the generic supergravity solution. We will generally work in the center of mass frame, where $m_{210}=m_{201}=0$, because it enormously simplifies the calculations. In the $M=1$ $AdS_5 \times S^5$ solution, only the $m_{2k00}$ momenta are non-zero. Later, it will be relevant that $m_{200} = L^4$ and $m_{400} = L^8/N^2 \, \text{dim} \left( \mathbb{L} \right)$ for a generic bubbling geometry.

We now expand all functions appearing in the metric at large $\rho$ and determine the asymptotic form of \eqref{eq:metricbubblingsurface} as a perturbation of \eqref{eq:metricbubblingvacuum}, expressed as a power series in $1/\rho$.

In the new coordinates, the flat half-space metric $ds^2_X$ is given by
\beq
\label{eq:metricX}
ds_X^2 = dy^2 + dx_i dx_i = \frac{\rho^2 + \sin^2 \theta}{\rho^2 + 1} \, d\rho^2 + \left( \rho^2 + \sin^2 \theta\right) d\theta^2 + \rho^2 \sin^2\theta \, d\alpha^2. 
\eeq
Using that $V$ is only defined up to an exact form we can find $\omega$ such that $V + d \omega$ does not have $\rho$ component at each order in $1/\rho$. This is useful because we want the coordinate $\rho $ to not mix with any of the other coordinates, in such a way that the metric is manifestly asymptotically $AdS_5 \times S^5$. Such $\omega$ was found in appendix D of \cite{Drukker:2008wr}.

We will only need the first few orders in the expansion: 
\beq 
\omega &=  -\frac{1}{2} \alpha  (M-1)+\frac{1}{2 L^2 \rho }\csc \theta (m_{001} \cos \alpha-m_{010} \sin \alpha)+\\
&+\frac{1}{4 L^4 \rho ^2}\csc ^2\theta ((m_{002}-m_{020}) \sin 2 \alpha+2 m_{011} \cos 2 \alpha)+ \\
&+\frac{\sin \theta}{6 L^6 \rho ^3} \Big(\csc ^4\theta  \Big(\left(3 m_{012}-m_{030}\right) \sin 3 \alpha-\left(m_{003}-3 m_{021}\right) \cos 3 \alpha\Big)+\\
&+2 m_{210} \sin \alpha-2 m_{201} \cos \alpha\Big)+\frac{\sin ^2\theta}{8 L^8 \rho ^4} \Big(\csc ^6\theta (4 (m_{031}-m_{013}) \cos 4 \alpha+ \\
&-(m_{004}-6 m_{022}+m_{040}) \sin 4 \alpha)+4 (m_{220}-m_{202}) \sin 2 \alpha-8 m_{211} \cos 2 \alpha\Big) + \CO (\rho^{-5}).
\eeq
This reduces to equation (A.3) of \cite{Gentle:2015ruo} or equation (D.1) of \cite{Drukker:2008wr} when imposing the center of mass condition.

Now we continue by changing coordinates in the metric \eqref{eq:metricbubblingsurface} and expanding at large $\rho$ to get an explicitly asymptotically $AdS_5 \times S^5$ solution: 
\beq
\label{eq:largeRhoMetric}
ds^2 &= \frac{1}{\left(\rho^2+1\right)} \left( 1+ F_\rho\right) d\rho^2+ \left(\rho^2+1\right) \left(1+F_1\right)\, ds^2_{AdS_3} + \rho^2\left( 1+ F_2\right) d\psi^2  \\
& + \cos^2\theta \left(1+F_3\right) \, ds^2_{S^3} + \left( 1+ F_4\right) d\theta^2 + \sin^2\theta \left( 1+ F_5\right) d\phi^2   \\
&  +F_6\, d\theta\, d\psi  +F_7\, d\psi\, d\phi +F_8\, d\theta\, d\phi.
\eeq
The first few orders of $F_\rho, F_1, \ldots, F_8$ in the center of mass frame are given in appendix \ref{sec:appendixallfunctions}.

The goal now is to do a change of coordinates from \eqref{eq:largeRhoMetric} to a Fefferman-Graham form in the $AdS_5$ part of the geometry (preserving the $AdS_3 \times S^3 \times S^1$ of the bubbling geometry ansatz): 
\beq 
\label{eq:FGform}
ds^2 &=  \frac{1}{u^2} \left( du^2 + \alpha_1 ds^2_{AdS_3} + \alpha_2 d\tilde{\psi}^2 \right) 
+ \alpha_3 ds^2_{S^3} + \alpha_4 d\tilde{\theta}^2 + \alpha_5 d\tilde{\phi}^2 +\\
&+ \alpha_6 d\tilde{\theta} d\tilde{\psi} + \alpha_7 d\tilde{\psi} d\tilde{\phi} 
+ \alpha_8 d\tilde{\theta} d\tilde{\phi},
\eeq
where there are no cross terms between $du$ and any of the other coordinates. The condition of asymptoting to $AdS_5 \times S^5$ (with $AdS_3 \times S^1$ slicing) at leading order imposes the following behavior in the metric elements
\beq 
&\alpha_1 = 1 + \sum_{n=1}^\infty \alpha_1^{(n)}(\tilde \theta, \tilde \phi, \tilde \psi ) \; u^n, \quad \alpha_2 = 1 + \sum_{n=1}^\infty \alpha_2^{(n)}(\tilde \theta, \tilde \phi, \tilde \psi )\; u^n, \\ &\alpha_3 = \cos^2\tilde \theta + \sum_{n=1}^\infty \alpha_3^{(n)}(\tilde \theta, \tilde \phi, \tilde \psi ) \; u^n, \quad
\alpha_4 = 1 + \sum_{n=1}^\infty \alpha_4^{(n)}(\tilde \theta, \tilde \phi, \tilde \psi )\; u^n, , \\
&\alpha_5 = \sin^2\tilde \theta + \sum_{n=1}^\infty \alpha_5^{(n)}(\tilde \theta, \tilde \phi, \tilde \psi ) \; u^n, \quad \alpha_6= 0 + \sum_{n=1}^\infty \alpha_6^{(n)}(\tilde \theta, \tilde \phi, \tilde \psi ), \\
&\alpha_7= 0 + \sum_{n=1}^\infty \alpha_7^{(n)}(\tilde \theta, \tilde \phi, \tilde \psi ),\quad \alpha_8= 0 + \sum_{n=1}^\infty \alpha_8^{(n)}(\tilde \theta, \tilde \phi, \tilde \psi ),
\eeq

We can read off the leading term of the change of coordinates in a similar way:
\beq 
&\rho = \frac{1}{u}\left( 1 + \sum_{n=1}^\infty \rho^{(n)}(\tilde \theta, \tilde \phi, \tilde \psi ) \; u^n \right), \quad \psi = \tilde \psi + \sum_{n=1}^\infty \psi^{(n)}(\tilde \theta, \tilde \phi, \tilde \psi )\; u^n, \\ 
&\theta = \tilde \theta + \sum_{n=1}^\infty \theta^{(n)}(\tilde \theta, \tilde \phi, \tilde \psi ) \; u^n, \quad
\phi = \tilde \phi + \sum_{n=1}^\infty \phi^{(n)}(\tilde \theta, \tilde \phi, \tilde \psi )\; u^n, , \\
\eeq

Now we solve for the change of coordinates in the large $\rho$ (or small $u$) limit. We do that by inserting the previous expansion into \eqref{eq:largeRhoMetric} and equating to \eqref{eq:FGform} at each order in $u$. We find $\{\rho,\theta,\phi,\psi\}$ up to $\CO (u^4)$ in terms of the new coordinates (note that the functions $F_\rho, F_1, \ldots$ are evaluated at $(\tilde \theta,\tilde \phi, \tilde \psi)$):

\beq 
\rho(u,\tilde \theta,\tilde \phi, \tilde \psi) &= \frac{1}{u} + \frac{(F_\rho^{(2)}-1)}{4} u^2  + \frac{F_\rho^{(3)}}{6} u^3  + \\
&+\frac{1}{128} \left(-16 (F_\rho^{(2)})^2+16 F_\rho^{(2)}+3 \left(\d_{\tilde \theta} F_\rho^{(2)} \right)^2+16 F_\rho^{(4)}+3 \csc ^2\tilde \theta \left(\d_{\tilde \phi} F_\rho^{(2)}\right)^2\right)u^4  + \CO (u^5), \\
\theta(u,\tilde \theta,\tilde \phi, \tilde \psi) &= \tilde \theta + \frac{\d_{\tilde \theta}F_\rho^{(2)}}{8}u^2 +  \frac{\d_{\tilde \theta}F_\rho^{(3)}}{18}u^3 + \frac{1}{256} \left(\d_{\tilde \theta} F_{\rho }^{(2)} \left(\d_{\tilde \theta \tilde \theta} F_{\rho }^{(2)}-16 F_4^{(2)}-12 F_{\rho }^2+4\right)+ \right. \\
&\left. \quad \quad  +8 \d_{\tilde \theta} F_{\rho }^{(4)}+\csc ^2\tilde{\theta } \d_{\tilde \phi} F_{\rho }^{(4)} \left(\d_{\tilde \theta}\d_{\tilde \phi} F_{\rho }^{(4)}+\cot \tilde{\theta } \d_{\tilde \phi} F_{\rho }^{(4)}\right)\right)u^4 + \CO(u^5), \\
\phi(u,\tilde \theta,\tilde \phi, \tilde \psi) &= \tilde \phi + \frac{\d_{\tilde \phi}F_\rho^{(2)}}{8 \sin^2 \tilde \theta }u^2 + \frac{\d_{\tilde \phi}F_\rho^{(3)}}{18 \sin^2 \tilde \theta} u^3 + \frac{1}{256\sin^2 \tilde \theta} \Bigg( 8 \d_{\tilde \phi} F_{\rho }^{(4)} +\d_{\tilde \theta} F_{\rho }^{(2)} \d_{\tilde \theta \tilde \phi} F_{\rho }^{(2)}+\\
&\quad \quad + \d_{\tilde \phi} F_{\rho }^{(2)} \left(-4 \cot\tilde{\theta } \d_{\tilde \theta} F_{\rho }^{(2)}+\csc ^2 \tilde{\theta } \d_{\tilde \phi \tilde \phi} F_{\rho }^{(2)}-12 F_{\rho }^{(2)}-16 F_5^{(2)}+4\right)\Bigg) + \CO(u^5), \\
\psi(u,\tilde \theta,\tilde \phi, \tilde \psi) &= \tilde \psi + \frac{\d_{\psi}F_\rho^{(2)}}{16}u^4 + \frac{\d_{\psi}F_\rho^{(3)}}{30}u^5 + \CO(u^6). 
\eeq

However, in order to define the FG cut-off we need to find $\{u,\tilde \theta,\tilde \phi, \tilde \psi\}$ in terms of the old coordinates. This can be done by inverting the previous equations for large $\rho$ \cite{Gentle:2015ruo}:

\beq 
\label{eq:changeofcoordsfinal}
u &= \frac{1}{\rho} \left( 1  + \frac{F_\rho^{(2)}-1}{4\rho^2} + \frac{F_\rho^{(3)}}{6\rho^3} + \frac{16(F_\rho^{(4)}-F_\rho^{(2)}+1)-(\partial_\theta F_\rho^{(2)})^2- (\partial_\phi F_\rho^{(2)})^2 \csc^2\theta  }{128\rho^4} + \CO\left(\rho^{-5}\right) \right) \\
\tilde\psi &= \psi  - \frac{\partial_\psi F_\rho^{(2)}}{16\rho^4} - \frac{\partial_\psi F_\rho^{(3)}}{30\rho^5}+ \CO\left(\rho^{-6}\right)  \\
\tilde\theta &= \theta  - \frac{\partial_\theta F_\rho^{(2)}}{8\rho^2} - \frac{\partial_\theta F_\rho^{(3)}}{18\rho^3} + \frac{1}{256\rho^4} \left[ -8 \partial_\theta F_\rho^{(4)} + 3 \partial_\phi F_\rho^{(2)}\, \partial_\theta \partial_\phi F_\rho^{(2)} \csc^2\theta \right.  \\
&\phantom{=\ }\left.- (\partial_\phi F_\rho^{(2)})^2 \cot\theta \csc^2\theta  +\partial_\theta F_\rho^{(2)} \left( 12-4 F_\rho^{(2)} +16 F_4^{(2)} +3 \partial_\theta^2 F_\rho^{(2)}  \right) \right]+\CO\left(\rho^{-5}\right)  \\
\tilde\phi &= \phi  - \frac{\partial_\phi F_\rho^{(2)}}{8\sin^2\theta\, \rho^2} - \frac{\partial_\phi F_\rho^{(3)}}{18\sin^2\theta\, \rho^3} + \frac{1}{256\sin^2\theta\, \rho^4}  \left[ -8\partial_\phi F_\rho^{(4)}+3 \partial_\theta F_\rho^{(2)}\, \partial_\theta\partial_\phi F_\rho^{(2)} \right.  \\
&\phantom{=\ }\left. + \partial_\phi F_\rho^{(2)} \left(12-4 F_\rho^{(2)} +16 F_5^{(2)} +3 \partial_\phi^2 F_\rho^{(2)} \csc^2\theta - 4 \partial_\theta F_\rho^{(2)} \cot\theta \right) \right]+ \CO \left(\rho^{-5}\right). 
\eeq

The cut-off is found by fixing $u = \epsilon$ and inverting order by order for small $\epsilon$ the first equation in \eqref{eq:changeofcoordsfinal} \cite{Jensen:2013lxa}: 
\beq
\label{eq:cutoff}
\rho_{cut-off}&(\epsilon,\psi,\theta,\phi) =  \frac{1}{\epsilon} + \frac{F_\rho^{(2)}-1}{4}\,\epsilon +  \frac{F_\rho^{(3)}}{6}\, \epsilon^2     \\
& + \frac{16\left(F_\rho^{(4)}-F_\rho^{(2)}\left(F_\rho^{(2)}-1\right)\right)\left(\partial_\theta F_\rho^{(2)}\right)^2- \left(\partial_\phi F_\rho^{(2)}\right)^2 \csc^2\theta}{128}\,  \epsilon^3 + \CO\left(\epsilon^4\right)= \\
&= \frac{1}{\epsilon} + {\epsilon \over 16}  \Big(12 \sin ^2\theta \left(2 \sin 2 \alpha m_{211}+\cos 2 \alpha \left(m_{220}-m_{202}\right)\right)+\\
&-3 \cos 2 \theta \left(2 m_{202}+2
   m_{220}-m_{400}+1\right)+2 m_{202}+2 m_{220}-m_{400}-3\Big)+\\
&+{\epsilon^2 \over 2 } \Big( 3\sin \alpha (\sin \theta-\sin 3 \theta) \left(m_{203}+m_{221}-m_{401}\right)+\\
&+4 \sin ^3\theta \left(\cos 3 \alpha \left(m_{230}-3 m_{212}\right)-\sin 3 \alpha
   \left(m_{203}-3 m_{221}\right)\right)+\\
&+3 \cos \alpha (\sin \theta-\sin 3 \theta) \left(m_{212}+m_{230}-m_{410}\right)\Big)+\\
&+ \epsilon^3 \Big( 40 \sin 4 \alpha \left(-8 m_{213}+9 m_{211} \left(m_{202}-m_{220}\right)+8 m_{231}\right) \sin ^4\theta+\\
&-10 \cos 4 \alpha \left(9 \left(m_{202}-m_{220}\right){}^2-4 \left(9
   m_{211}^2+2 m_{204}+2 \left(m_{240}-6 m_{222}\right)\right)\right) \sin ^4\theta+\\
   &+4 \sin 2 \alpha \sin ^2(\theta
   ) \Big(-16 \left(m_{213}+m_{231}\right)-3 m_{211} \left(18 m_{202}+18 m_{220}-9
   m_{400}+1\right)+\\
   &+24 m_{411}+\cos 2 \theta \left(-80 \left(m_{213}+m_{231}\right)+m_{211} \left(90 m_{202}+90 m_{220}-45 m_{400}-3\right)+120 m_{411}\right)\Big) +\\
   &-2 \cos 2 \alpha \sin ^2\theta \Big(-54 m_{202}^2+3 \left(9 m_{400}-1\right) m_{202}-16 m_{204}+\\
   &+3 m_{220} \left(18 m_{220}-9 m_{400}+1\right)+8 \left(2 m_{240}+3 m_{402}-3
   m_{420}\right)+\\
   &+\cos 2 \theta \big(90 m_{202}^2-3 \left(15
   m_{400}+1\right) m_{202}-80 m_{204}+3 m_{220} \left(-30 m_{220}+15 m_{400}+1\right)+\\
   &+40 \left(2 m_{240}+3 m_{402}-3 m_{420}\right)\big)\Big) +\\
   &-\frac{3}{8} \Big(238 m_{202}^2-4 \left(19 m_{220}+25 m_{400}-9\right) m_{202}+552 m_{211}^2+238 m_{220}^2+\\
   &-48 m_{204}-96 m_{222}-48 m_{240}+4
   m_{220} \left(9-25 m_{400}\right)+m_{400} \left(25 m_{400}-18\right)+\\
   &+96 m_{402}+96 m_{420}+9\Big)+\\
   &-\frac{1}{8} \cos 4 \theta \big(270 m_{202}^2+12 \left(15 m_{220}-15
   m_{400}-1\right) m_{202}+360 m_{211}^2+270 m_{220}^2+\\
   &+45 m_{400}^2-240 m_{204}-480 m_{222}-240 m_{240}+6 m_{400}-12 m_{220} \left(15 m_{400}+1\right)+\\
   &+480 m_{402}+480 m_{420}-80
   m_{600}+29\big)+\\
   &+\cos 2 \theta \big(99 m_{202}^2-6 \left(9 m_{220}+6 m_{400}+2\right) m_{202}+252 m_{211}^2+99 m_{220}^2+9 m_{400}^2+\\
   &-24 m_{204}-48 m_{222}-24 m_{240}+6
   m_{400}-12 m_{220} \left(3 m_{400}+1\right)+\\
   &+48 m_{402}+48 m_{420}-8 m_{600}-7\big)+6 m_{600} \Big) + \CO\left(\epsilon^4\right)
\eeq

This cut-off defines a nine-dimensional hypersurface as a regulated boundary of the ten-dimensional geometry.

To finish this section, we discuss the choice of cut-off surface and compare it with other choices in the literature, in particular the one used by Skenderis and Taylor when applying their Kaluza-Klein holography program \cite{Skenderis:2006uy} to the Lin-Lunin-Maldacena bubbling geometries \cite{Skenderis:2007yb}. Note that the bubbling geometries we study in this work, dual to surface operators, are a double analytic continuation of those \cite{Gomis:2007fi}, and formulas from \cite{Skenderis:2007yb} can be used to compute the expectation value of some operators in the presence of the surface defect (see section 4 in \cite{Drukker:2008wr}) such as $\langle \CO_\Sigma \; \CO_{\Delta, k} (x) \rangle$, where $\CO_{\Delta, k} (x)$ are chiral primary operators, $\langle \CO_\Sigma \; T_{\mu \nu} (x) \rangle$ or $\langle \CO_\Sigma W_C \rangle$, where $W_C$ is a Wilson loop operator. Note that this means that $c_2$ has been computed holographically for a generic $\CO_\Sigma$ thanks to the Kaluza-Klein holography method. 

The key idea in Kaluza-Klein holography is that given any asymptotically $AdS_5 \times S^5$ solution of type IIB supergravity, even when it is not an uplift of a solution of $AdS_5$ (super)gravity, one can systematically compute expectation values of light probes in that background. This is achieved by systematically Kaluza-Klein reducing the solution on the asymptotic $S^5$ to a solution of a theory of gravity in $AdS_5$ coupled to an infinite number of fields. The five-dimensional action is constructed as follows:
\begin{enumerate}
    \item Express the full $AdS_5 \times S^5$ solution as a perturbation of the $AdS_5 \times S^5$ vacuum: $\Psi = \Psi_0 + \psi$, where $\Psi$ denotes all the fields of type IIB supergravity. 
    \item Expand the perturbation in $S^5$ spherical harmonics $\psi = \psi^I \, Y^I$, where $Y^I$ collectively denotes scalar, vector, tensor etc. $SO(6)$ spherical harmonics. 
    \item Find the equations of motion of $\psi^I$ derived from type IIB supergravity equations of motion\footnote{There is a subtlety related to gauge invariance: since the ten-dimensional theory must be diffeomorphism invariant, not all fluctuations are independent. In \cite{Skenderis:2006uy,Skenderis:2007yb} the authors find the gauge invariant fluctuations, instead of fixing the gauge as it was previously done in the literature. The gauge invariant fluctuation must be found order by order in the number of fields.} and expand them perturbatively in the number of fields: 
    $$
    \CL_I \psi^I  =  \CL_{IJ} \psi^I\psi^J +\CL_{IJK} \psi^I\psi^J \psi^K + \ldots,
    $$
    where the $\CL$ denote differential operators. This equation cannot be integrated into a five-dimensional action, since there are derivative couplings in the right hand side in the nonlinear couplings. 
    \item Diagonalize the previous equations at the linear order, we denote them by $s^I$, and find non-linear transformations 
    $$
    \hat \Psi^I = s^I + \CK^I_{JL} s^J\, s^K + \ldots
    $$
    such that the equations of motion of $\hat \Psi^I$ can be integrated into a five-dimensional action. This is what is called the non-linear Kaluza-Klein map.
\end{enumerate}
The 5d action is systematically constructed perturbatively in the number of fluctuations. Using the holographic dictionary of \cite{Witten:1998qj,Gubser:1998bc} we know that to compute an expectation value of a probe of dimension $\Delta$, and only a finite number fluctuations will be needed. The correlation function can then be computed using holographic renormalization \cite{Skenderis:2002wp}. 

It is after the Kaluza-Klein map has been done that the choice of regularization and Fefferman-Graham cut-off appears. We will compare our choice with their choice in \cite{Skenderis:2007yb}, where they extract the expectation values from the asymptotics of the fields. They define a ten-dimensional Fefferman-Graham metric and a five-dimensional Fefferman-Graham metric. We want to stress that our Fefferman-Graham cut-off  \eqref{eq:cutoff} is not equivalent to $z = \epsilon$ where $z$ is the FG coordinate of 2.30 in \cite{Skenderis:2007yb}. That is why their change of coordinates in equation 3.53 is different from our change of coordinates \eqref{eq:changeofcoordsfinal}, even though in both cases we claim to have defined an $AdS_5$ radial coordinate. The difference is that their definition of 10d metric in equation 2.29 restricts to singlets of the $SO(6)$ spherical harmonics. One can go from our cut-off to the their cut-off by projecting out all contributions from non- $SO(6)$-invariant spherical harmonics.

\section{On-shell action}
\label{sec:onshellactionsec}
In this section, we review the evaluation of the supergravity action in the generic bubbling geometry solution \cite{Gentle:2015ruo}. Since only the 5-form field strength and the metric are nonzero in this solution, the only terms in the action are 
\beq
S = \frac{1}{2 \kappa^2} \left( \int d^{10}x \, \sqrt{g} R - 4 \int F_{(5)} \wedge \star F_{(5)}  + 2 \int d^9x \sqrt{h} \, K \right),
\eeq
where $\kappa^2 = 8 \pi G_N^{(10d)}$. 

This bulk pseudo-action evaluates to zero since $R=0$ and $F_5$ is self dual. However, one can use the prescription from \cite{Giddings:2001yu,DeWolfe:2002nn} (which is the same as \cite{Kurlyand:2022vzv} and was generalized in \cite{Frey:2019fqz}). We can split $F_5$ into an electric and a magnetic part such that $\star F_{el} = F_{mag}$, and we consider the contribution of the kinetic term of $F_{(5)}$ to the action to be
\beq 
-\frac{1}{2 \kappa^2} \left( \int_{10d} \, 2  4 \,  F_{(5)}^{(el)} \wedge \star F_{(5)}^{(el)} \right),
\eeq
where
\beq 
\label{eq:F_5bubblingsurfacesplitting}
F_{(5)}^{(el)} &=- \frac{1}{4} \left[ d \left( y^2 \frac{2z + 1}{2 z -1} (d\chi +V) \right) - y^3 \star_X d \left( \frac{1}{y^2} \left(z+\frac{1}{2} \right) \right) \right] \wedge {\vol}_{AdS_3}, \\
\star F_{(5)}^{(el)}  &=F_{(5)}^{(mag)} = -\frac{1}{4} \left[ d \left( y^2 \frac{2z - 1}{2 z +1} (d\chi +V) \right) - y^3 \star_X d \left( \frac{1}{y^2} \left(z-\frac{1}{2} \right) \right) \right] \wedge {\vol}_{S^3}.
\eeq

Using the equation of motion of $V$ \eqref{eq:difeqforV} we find 
\beq 
&F_5^{(el)} \wedge \star F_5^{(el)} = -
\frac{y z }{2(1-4z^2)^2} \left( (1 - 4z^2)^2 + \frac{2y}{z}(1 - 4z^2) \d_y z + 4y^2 \d_I z\,\d^I z \right) \times \\
&\times{\vol}_{AdS_3} \wedge {\vol}_{S^3} \wedge d \chi \wedge {\vol}_{X}= \\
&=\left( -\frac{1}{2} y z + \d_I u_I + \frac{y^3}{4(1-4z^2)}\left( y \; d \left( \frac{1}{y} \star_X dz \right) \right) \right) \times {\vol}_{AdS_3} \wedge {\vol}_{S^3} \wedge d \chi \wedge {\vol}_{X},
\eeq
where 
\beq 
\label{eq:ui}
u_I = - \frac{y^3}{4(1-4z^2)} \, \d_I z , \quad I= x_1,\,x_2,\, y. 
\eeq

The last term vanishes because $d(dV)$ is proportional to delta functions supported on the sources but the coefficient vanishes. Then, the on-shell action will have contributions from three integrals: 
\beq \label{eq:integrals}
I_1=& - { 4 \over \kappa^2 } \vol (AdS_3) \; \vol (S^3) \; \vol (S^1) \;  \int_X -\frac{1}{2} y \, z \, dx_1\, dx_2 \, dy, \\
I_2=&  - { 4 \over \kappa^2 } \vol (AdS_3) \; \vol (S^3) \; \vol (S^1) \; \int_X \d_I u^I dx_1\, dx_2 \, dy, \\
I_3=& {1 \over \kappa^2} \int d^9x \sqrt{h} \, K .
\eeq
All these integrals are divergent and have to be regularized using the defect Fefferman Graham cut-off \eqref{eq:cutoff}. They were computed in appendix E of \cite{Gentle:2015ruo}, we reproduce those calculations for completeness in appendix \ref{appendixintegrals}. 

The final result is 
\beq 
S_{on-shell} = \frac{\pi}{2\kappa^2} \text{Vol}(AdS_3) \, \text{Vol}(S^3) \, \text{Vol}(S^1) \left( \frac{5}{\epsilon^4} + \frac{2}{\epsilon^2} + \frac{3}{8} - m_{400} - \mathcal{F} \right),
\eeq
where 
\beq 
\label{eq:calF}
\mathcal{F} &= \frac{3}{32} \left( 1 +4 m_{220}  +4 m_{202}-2m_{400} +10\left(m_{220}^2 +m_{202}^2\right)+\right. \\
&\left.+24 m_{211}^2-4\left(m_{220}+m_{202}\right)m_{400}+m_{400}^2 -4 m_{220} m_{202} \right).
\eeq

\section{Counterterm}
\label{sec:counterterm}
Now we will show that if we add a cosmological constant counterterm in the regulated nine-dimensional boundary
\beq 
\label{eq:countertermlab}
 \frac{a}{2\kappa^2\, L} \int d^9x \sqrt{h},
\eeq
we will be able to cancel the term $\mathcal{F}$. We have added a factor of the $AdS_5$ length scale by dimensional analysis. The calculation is in appendix \ref{appendixarea}.

The final result is 
\beq 
\frac{a}{2\kappa^2\, L} \int d^9x \sqrt{h} = \frac{a}{2\kappa^2\, L} \vol (AdS_3) \, \vol (S^3) \, \vol (S^1) \,  \pi \left( \frac{\epsilon^4}{2} + \frac{\epsilon^2}{4} + \mathcal{C} \right),
\eeq
where 
\beq 
\mathcal{C} &= \frac{1}{32} \left(-10 m_{2,0,2}^2+4 \left(m_{2,2,0}+m_{4,0,0}-1\right) m_{2,0,2}-24 m_{2,1,1}^2-10 m_{2,2,0}^2-m_{4,0,0}^2+ \right. \\
&\left.-4 m_{2,2,0}+4 m_{2,2,0} m_{4,0,0}+2 m_{4,0,0}-1\right)
\eeq

Then, fixing the coefficient of the counterterm to $a = -3$ we find that the total on-shell action is given by 
\beq 
S_{on-shell} + S_{counterterm} = \frac{\pi}{2\kappa^2} \text{Vol}(AdS_3) \, \text{Vol}(S^3) \, \text{Vol}(S^1) \left( \frac{7}{2\epsilon^4} + \frac{5}{4\epsilon^2} + \frac{3}{8} - m_{400}  \right),
\eeq

In order to cancel the power-law divergencies we use background subtraction: we compute the difference of the generic bubbling geometry on-shell action and the $AdS_5 \times S^5$ background. Note that there is no 4d conformal anomaly because $\CN=4$ super Yang-Mills lives in the conformal boundary of these geometries, which is always $AdS_3 \times S^1$.  

The regularized on-shell action after including the cosmological constant counterterm is 
\beq 
\label{eq:final}
S_{on-shell}- S_{on-shell}^{\text{vacuum}} &=\frac{\pi}{2\kappa^2} L^8\, \text{Vol}(AdS_3) \, \text{Vol}(S^3) \, \text{Vol}(S^1) \left( 1 - L^{-8} m_{400}  \right) = \\
&=\log \Lambda \; \left( N^2 - \sum_l N_l^2 \right),
\eeq
where we have used that 
\beq 
2 \kappa^2 = (2 \pi)^7 g_s^2 (\alpha')^4, \quad L^4 = 4 \pi g_s (\alpha')^2 N, \quad \vol (AdS_3) = 2 \pi \log \Lambda,
\eeq
in particular that the regularized volume of $AdS_3$ with global coordinates gives a logarithmic divergence. Note that the dual of $\CO_{\mathbb R^2}$ would correspond to $AdS_3$ in the Poincar\'e patch and does not have any conformal anomaly. The result \eqref{eq:final} reproduces exactly the b-anomaly of the surface operator in equation \eqref{eq:banomaly}.

\section{Conclusions}
\label{conclusions}

We have successfully reproduced the Euler anomaly $b$ of the $1/2$ BPS surface operator of $\CN = 4$ super Yang-Mills from the on-shell type IIB supergravity action. To do this, we have proposed a 9d boundary counterterm to be added to the 10d supergravity action. This is a first step towards renormalization from a higher dimensional perspective, and we have not developed a microscopic interpretation of the counterterm. The renormalization prescription should not merely cancel divergences and match results, but should admit a deeper interpretation within the holographic correspondence, in the spirit of \cite{Papadimitriou:2004ap}.



Our prescription produces the correct Euler anomaly after subtracting the vacuum on-shell action. It is crucial that the $AdS_5 \times S^5$ background is in the coordinates \eqref{eq:metricbubblingvacuum}, that make the symmetries of the surface operator manifest, and that the vacuum on-shell action $S_{\text{on-shell}}^0$ is regulated with the cut-off \eqref{eq:cutoff} (which is not $\rho = 1 /\epsilon$). The natural next step is to find which covariant counterterms in the 9d regulated boundary can cancel these divergencies while still keeping the same finite term $\CO (\epsilon^0)$, similarly to \cite{Emparan:1999pm}. Studying different cut-off choices could give new insights into the physics of holographic renormalization in the presence of defects. Our result suggests that the prescription of \cite{Papadimitriou:2004rz,Estes:2012nx,Jensen:2013lxa}, which has been used successfully in the context of Ryu-Takayanagi surfaces in the presence of a defect, can also be used to compute conformal anomalies and certain physical terms in the on-shell action. 

On the field theory side, it would be interesting to understand if $b$ can be extracted from a correlation function of local operators in the presence of the surface operator\footnote{There is some recent work relating $b$ and the four-point function of displacement operators \cite{drukker_seminar_2025}. However, the holographic dual of the displacement operator remains unknown, even within the standard $AdS_{d+1}/CFT_d$ dictionary. See \cite{Bachas:2024nvh} for a proposal. To our knowledge, how to approach this problem in bubbling geometries is an open problem.}, in analogy with how the central charge is obtained in 2d from the partition function or the two-point function of the stress tensor. If such correlation functions were accessible through the \textit{Kaluza-Klein holography} method, they could provide further understanding of higher dimensional holography. Similarly, one could try to exploit the relationship between $d_2$, $b$ and entanglement entropy \cite{Jensen:2018rxu} to gain insight into the structure of the allowed 9d counterterms. 

Several backreacted geometries dual to supersymmetric defects are known \cite{Kraus:1998hv,Lin:2004nb,DHoker:2007mci,Gomis:2007fi,Yamaguchi:2006te,Gomis:2006cu,DHoker:2007hhe,Gaiotto:2009gz}, many of which have some analogue of the $b$ anomaly of the surface operator (conformal anomalies or contributions to the free energy)\footnote{This idea does not restrict to defect CFTs. For example, in \cite{Gutperle:2017tjo}, the authors holographically compute the 10d on-shell action and the entanglement entropy of a spherical region via the Ryu–Takayanagi prescription for certain $AdS_6\times S^2 \times \Sigma_2$ solutions of type IIA supergravity. In that case, the cutoff prescription is clearer, and no additional boundary terms are required. Understanding this example and the comparison with other bubbling geometries in greater detail would be very valuable. Other works discussing 10d/11d on-shell actions without defects include \cite{Kurlyand:2022vzv,Beccaria:2023hhi}.}. Whether the renormalization prescription proposed here applies to these cases remains an important open direction, which we are currently exploring in ongoing work. 

Recent progress applying equivariant localization to supergravity \cite{BenettiGenolini:2023kxp} proposes another interesting new direction. This method has already been used for precision tests of holography, such as computing central charges, on-shell actions, free energy, and different charges of several gauged supergravity solutions \cite{BenettiGenolini:2024lbj,BenettiGenolini:2024xeo,BenettiGenolini:2023yfe,BenettiGenolini:2023ndb}. Extending these methods to bubbling geometries in ten and eleven dimensions would be a valuable development.
 
One of the main conceptual challenges in understanding holographic renormalization in higher dimensions is dealing with a degenerate boundary. The conformal boundary of the 10d asymptotically $AdS_5 \times S^5$ spacetime is four dimensional, and there is where the CFT ``lives''. However, the physical interpretation of the regulated 9d cut-off surface is not clear, and it is also not known how to fix the boundary conditions and boundary terms. In the standard AdS/CFT correspondence, varying the Fefferman–Graham cutoff $\epsilon$ has a well understood interpretation in terms of the energy scale of the CFT and holographic RG flows. Extending this understanding to ten dimensions remains an important open problem. Studying precise holographic models from a higher-dimensional perspective can help us better understand the holographic nature of the duality and the emergence of geometry from the boundary quantum theory.

\section*{Acknowledgments}

I am extremely grateful to my supervisor, Jaume Gomis, for suggesting this problem, for many essential discussions on the topic, and for carefully reading and commenting on the draft. I would also like to thank  A. Frey, J. Gauntlett, M. Gutperle, S. Komatsu, J. Maldacena, and A. O'Bannon for insightful discussions. I am also very grateful to D. Rodr\'iguez-G\'omez and J. Munday for their careful reading and constructive comments on the draft.  I would also like to thank the organizers of the program \textit{Quantum field theory with boundaries, impurities, and defects} at the Isaac Newton Institute for Mathematical Sciences, Cambridge, where the results of this work were presented. I also thank the INI for support and hospitality. This work was supported by EPSRC grant no EP/Z000580/1. Research at Perimeter Institute is supported in part by the Government of Canada through
the Department of Innovation, Science and Economic Development Canada and by the
Province of Ontario through the Ministry of Colleges and Universities. The project that gave rise to these results
received the support of a fellowship from ``la
Caixa'' Foundation (ID 100010434). The
fellowship code is LCF/BQ/EU23/12010100.

\begin{appendix}

\section{Large \texorpdfstring{$\rho$}{rho} expansion of the metric functions}
\label{sec:appendixallfunctions}

Here we write the first few orders of the functions $F_\rho, F_I$ with $I= 1,\ldots,8$  in \eqref{eq:largeRhoMetric}. These functions can be computed as an asymptotic series in $\rho$:
\beq 
F_a (\rho, \theta, \alpha) = \sum_{n=1}^\infty  \frac{F_a^{(n)}\left(\theta, \alpha\right)}{\rho^n},
\eeq
where $a= \rho,1 \ldots,8$. We will gauge fix the position of the center of mass to zero, which translates into $m_{201}=m_{210}=0$, this makes the expressions much simpler. we have fixed $L=1$, but that can be easily restored using the dimensions of $m_{abc}$.

\beq 
4  F_{\rho}^{(2)} &=  \left(1-3\cos 2\theta\right) \left[ 1 + 2 \left(m_{220} +m_{202}\right) - m_{400}   \right] + \\
&  +12\left[ \cos 2\alpha \left(m_{220} - m_{202} \right) +2 m_{211}\sin 2\alpha \right] \sin^2\theta\\
 F_{\rho}^{(3)} &= 3 \left(\sin\theta-\sin 3\theta  \right)  \left[   \left( m_{212} + m_{230} - m_{410} \right)\cos\alpha + \left( m_{221} + m_{203} - m_{401} \right)\sin\alpha  \right] +\\
\;&+4 \sin^3\theta \left[ \left( -3 m_{212} + m_{230} \right)\cos 3\alpha -  \left(-3 m_{221} + m_{203}\right)\sin 3\alpha  \right] \\
32  F_{\rho}^{(4)} &= -4  \cos^4\theta +\left(5-12\cos 2\theta +15 \cos 4\theta\right) \left(2m_{202}+2m_{220}-m_{400}\right) +\\
&-16 \left(1+5\cos 2\theta\right) \sin^2\theta \sin 2\alpha \left[ 3 m_{211} +8 \left( m_{213}+m_{231}\right) -12m_{411} \right]+\\
&-8 \left(1+5\cos 2\theta\right) \sin^2\theta \cos 2\alpha \left[ 3 \left( m_{220}-m_{202}\right)+8 \left(m_{240}-m_{204}\right)+12\left(m_{402}-m_{420}\right)\right]+\\
& -640   \sin 4\alpha \sin^4\theta \left( m_{213}-m_{231}\right)+24\left(3-4\cos 2\theta +5\cos 4\theta -40\cos 4\alpha\sin^4 \theta \right) m_{222}+\\
& +4 \left( 9-12\cos 2\theta+15\cos 4\theta+40\cos 4\alpha \sin^4 \theta \right) \left( m_{204}+m_{240}\right)\\
& - 4  \left(3-4\cos 2\theta+5\cos 4\theta\right) \left[6\left(m_{402}+m_{420}\right) -m_{600}\right]+\\
&-\Big(12 \sin^2 \theta \left[\cos 2\alpha \left( m_{202}-m_{220}\right) -2\sin 2\alpha \;m_{211}\right]  +\\
& -(1-3\cos 2\theta) \left(2m_{202}+2m_{220}-m_{400}\right)\Big)^2
\eeq

\beq
 F_1^{(2)}&= 3 \sin ^2\theta \left(\cos 2 \alpha \left(m_{202}-m_{220}\right)-2 \sin 2 \alpha m_{211}\right)+\\
 &+\frac{1}{4} (3 \cos 2 \theta-1) \left(2 m_{202}+2 m_{220}-m_{400}+1\right),\\
F_1^{(3)}&=3 \sin \alpha (\sin 3 \theta-\sin \theta) \left(m_{203}+m_{221}-m_{401}\right)+\\
&+4 \sin ^3\theta \left(\sin 3 \alpha \left(m_{203}-3 m_{221}\right)+\cos 3 \alpha \left(3
m_{212}-m_{230}\right)\right)+\\
&+3 \cos \alpha (\sin 3 \theta-\sin \theta) \left(m_{212}+m_{230}-m_{410}\right), \\
\eeq

\beq
F_2^{(2)}&= 3 \sin ^2\theta \left(\cos 2 \alpha \left(m_{202}-m_{220}\right)-2 \sin 2 \alpha m_{211}\right)+\\
&+\frac{1}{4} (3 \cos 2 \theta-1) \left(2 m_{202}+2 m_{220}-m_{400}+1\right), \\
F_2^{(3)}&= 3 \sin \alpha (\sin 3 \theta-\sin \theta) \left(m_{203}+m_{221}-m_{401}\right)+\\
&+4 \sin ^3\theta \left(\sin 3 \alpha \left(m_{203}-3 m_{221}\right)+\cos 3 \alpha \left(3 m_{212}-m_{230}\right)\right)+\\
&+3 \cos \alpha (\sin 3 \theta-\sin \theta) \left(m_{212}+m_{230}-m_{410}\right) \\
\eeq

\beq
F_3^{(2)}&=  6 \sin 2 \alpha \sin ^2\theta m_{211}+3 \cos 2 \alpha \sin ^2\theta \left(m_{220}-m_{202}\right)-\\
&+\frac{1}{4} (3 \cos 2 \theta-1) \left(2 m_{202}+2 m_{220}-m_{400}+1\right) \\
F_3^{(3)}&= 3 \sin \alpha (\sin \theta-\sin 3 \theta) \left(m_{203}+m_{221}-m_{401}\right)+\\
&+4 \sin ^3\theta \left(\cos 3 \alpha \left(m_{230}-3 m_{212}\right)-\sin 3 \alpha \left(m_{203}-3 m_{221}\right)\right)+\\
&+3 \cos \alpha (\sin \theta-\sin 3 \theta) \left(m_{212}+m_{230}-m_{410}\right)  \\
\eeq

\beq
F_4^{(2)}&= 6 \sin 2 \alpha \sin ^2\theta m_{211}+3 \cos 2 \alpha
   \sin ^2\theta \left(m_{220}-m_{202}\right)+\\
   &-\frac{1}{4} (3 \cos (2
   \theta )-1) \left(2 m_{202}+2 m_{220}-m_{400}+1\right)  \\
F_4^{(3)}&=  \sin (\alpha
   ) (\sin \theta-\sin 3 \theta)
   \left(m_{203}+m_{221}-m_{401}\right)+\\
   &+4 \sin ^3\theta \left(\cos (3
   \alpha ) \left(m_{230}-3 m_{212}\right)-\sin 3 \alpha
   \left(m_{203}-3 m_{221}\right)\right)+\\
   &+3 \cos \alpha (\sin (\theta
   )-\sin 3 \theta) \left(m_{212}+m_{230}-m_{410}\right)  \\
\eeq

We omit $F_4^{(2)} = F_5^{(2)},F_4^{(3)} = F_5^{(3)}$, but note that $F_4^{(3)} \neq F_5^{(4)}$. 

\beq
F_6^{(2)}&= \cot \theta \csc ^4\theta \left(4 \cos 4 \alpha
   \left(m_{031}-m_{013}\right)-\sin 4 \alpha \left(m_{004}-6m_{022}+m_{040}\right)\right)+\\
   &-2 \sin 2 \theta \left(\sin 2 \alpha
    \left(m_{202}-m_{220}\right)+2 \cos 2 \alpha
   m_{211}\right)  \\
F_6^{(3)}&= 0  \\
\eeq

\beq
F_7^{(2)}&= \sin ^2\theta \Big(\csc ^6\theta \left(4 \sin 4 \alpha
   \left(m_{031}-m_{013}\right)+\cos 4 \alpha \left(m_{004}-6 m_{022}+m_{040}\right)\right)+\\
   &+8 \sin 2 \alpha m_{211}+4 \cos 2
   \alpha  \left(m_{220}-m_{202}\right)+4 m_{202}+4 m_{220}-2
   m_{400}+2\Big),  \\
F_7^{(3)}&=8 \sin ^3\theta \big(-3 \cos \alpha \cos 2
   \theta  \left(m_{212}+m_{230}-m_{410}\right)+\\
   &+2 \sin ^2\theta
   \left(\cos 3 \alpha \left(m_{230}-3 m_{212}\right)-\sin 3 \alpha
   \left(m_{203}-3 m_{221}\right)\right)+\\
   &-3 \sin \alpha \cos (2 \theta
   ) \left(m_{203}+m_{221}-m_{401}\right)\big)   \\
\eeq

\beq
F_8^{(4)}&= 4 \sin ^3\theta \cos \theta \left(\sin 2 \alpha
   \left(m_{202}-m_{220}\right)+2 \cos 2 \alpha m_{211}\right)+\\
   +&\cot
   \theta \csc ^2\theta \left(\sin 4 \alpha \left(m_{004}-6
   m_{022}+m_{040}\right)+4 \cos 4 \alpha
   \left(m_{013}-m_{031}\right)\right)  \\
\eeq

\section{Integrals}
\label{appendixintegrals}
In this appendix, we review the computation of the integrals \eqref{eq:integrals}, following the discussion in Section 4.2 and Appendix E of \cite{Gentle:2015ruo}.

We start $I_1$, which can be split into a sum of $M$ integrals that are easier to compute: 
\beq
I_1=- { 4 \over \kappa^2 } \vol (AdS_3) \; \vol (S^3) \; \vol (S^1) \;  \left({1 \over 2} \int_X \; y  \, dx_1 \, dx_2 \, dy + \sum_{l=1}^M \int_X \; y \, z_l \, dx_1 \, dx_2 \, dy\right),
\eeq
where $z_l$ is given in \eqref{eq:zandV}. The first integral can be computed immediately using the change of variables \eqref{changeofcoords} and the cut-off \eqref{eq:cutoff}:

\beq 
\label{eq:firstermI1}
&{1 \over 2} \int_X \; y  \, dx_1 \, dx_2 \, dy = {1 \over 2} \int_0^{2\pi}d \alpha \int_0^{\pi/2} d\theta \int_0^{\rho_{c}} d\rho \; \rho  \sin \theta \cos \theta \left(\sin ^2\theta+\rho ^2\right) = \\
&= {1 \over 2} \int_0^{2\pi}d \alpha \int_0^{\pi/2} d\theta \; \left[\frac{1}{8} \rho ^2 \sin \theta \cos \theta \left(\rho ^2+2 \sin ^2\theta\right)\right]_0^{\rho_{c}} = \\
&= \frac{\pi }{8 \epsilon ^4}+\frac{\pi }{32 \epsilon ^2} \left(1+2 m_{202}+2 m_{220}-m_{400} \right)+\\
&+\frac{\pi}{1536}  \Big(-288 m_{202}^2+144 m_{220} m_{202}+108 m_{400} m_{202}+12 m_{202}-720
   m_{211}^2-288 m_{220}^2+\\
   &-27 m_{400}^2+48 m_{204}+12 m_{220}+96 m_{222}+48 m_{240}+108 m_{220} m_{400}-6 m_{400}-96 m_{402}+\\
   &-96 m_{420}+16 m_{600}-7\Big)+O\left(\epsilon ^2\right),
\eeq
where we have used that the jacobian for the change of variables gives
\beq 
dx_1 \, dx_2 \, dy = \frac{\rho  \sin \theta \left(\sin ^2\theta+\rho ^2\right)}{\sqrt{\rho ^2+1}} \, d\rho \, d\theta \, d\alpha.
\eeq
Note that both the finite and divergent terms depend on the solution. 

To compute the remaining $M$ integrals it is useful to modify the change of coordinates \eqref{changeofcoordsgeneral} to be centered in each ``particle'' location:
\beq
\label{changeofcoordsxl}
\chi &= \frac{1}{2}( \bar \psi - \bar \phi), \quad \bar \alpha = \bar \psi + \bar \phi,\\
y &= y_l \, \sqrt{{\bar \rho}^2+1} \, \cos \bar \theta, \\
x_1 &= (\vec x_l)_1+ y_l \, \bar \rho \, \sin \bar \theta \cos \bar \alpha, \\
x_2 &=  (\vec x_l)_2+ y_l \, \bar \rho \, \sin \bar \theta \sin \bar \alpha, \\
\eeq
where $\bar \rho \in (0, \infty), \; \bar \theta \in \left[0, \pi/2\right] , \; \bar \psi \in \left[0, 2\pi\right)$ and $\bar \phi \in \[0, 2\pi\)$. The change of coordinates is different for each integral, we should denote them $\bar \rho_l, \bar \theta_l, \bar \alpha_l$ but we drop the index for readability. 

Using this, the integrals simplify
\beq 
&\int_X \; y \, z_l \, dx_1 \, dx_2 \, dy = y_l^4 \int \, d\bar \rho \; d\bar \alpha \; d\bar \theta \; \; \;  \bar \rho \; \cos^3 \bar \theta \; \sin \bar \theta =\\
&=   {y_l^4 \over 2 }\; \int \; d\bar \alpha \; d\bar \theta \; \; \;  \Big[\bar \rho^2 \; \cos^3 \bar \theta \; \sin \bar \theta \Big]_0^{\bar \rho_c}
\eeq

Note, however, that the integration limits will change accordingly. The angular variables ranges remain unchanged but we need to deduce a new FG cut-off, like we did in section \ref{sectioncutoff}. However, since the integral will be quadratic in $\bar \rho$, we only need to compute $\bar \rho_{cut-off}$ up to order $\CO(\epsilon)$ to compute the divergent and finite parts of the integral. We will compute this new cut-off from \eqref{eq:cutoffinalpha}. All these change of coordinates cannot be written in closed form but perturbatively in the radial coordinate. 

First, we need to express the old coordinates $\{\rho,\theta,\alpha\}$ in terms of the new  $\{\bar \rho, \bar \theta, \bar \alpha\}$ as a power series in $1/\bar \rho$: 
\beq 
\rho&=\bar{\rho } y_l+r_l \sin \bar{\theta } \cos \left(\bar{\alpha }+\beta _l\right)+\\
&+\frac{r_l^2 \left(\cos \left(2 \bar{\theta }\right)-2 \sin ^2\bar{\theta } \cos
   \left(2 \left(\bar{\alpha }+\beta _l\right)\right)+3\right)+4 \left(y_l^2-1\right) \cos ^2\bar{\theta }}{8 \bar{\rho } y_l}+\CO\left(\frac{1}{\bar{\rho }^2}\right),\\
\theta&=\bar{\theta }+\frac{r_l\cos \bar{\theta } \cos \left(\bar{\alpha }+\beta_l\right)}{\bar{\rho } y_l}-\frac{\cot \bar{\theta }}{2 \bar{\rho }^2 y_l^2} \Big(-\sin ^2\bar{\theta }+y_l^2 \sin ^2\bar{\theta }+\\
&+r_l^2 \left(\cos \left(2 \left(\bar{\alpha }+\beta_l\right)\right)-\cos \left(2 \bar{\theta }\right) \cos ^2\left(\bar{\alpha }+\beta_l
   \right)\right)\Big)+\CO\left(\frac{1}{\bar{\rho }^3}\right),\\
\alpha&=\bar{\alpha }-\frac{r_l\csc \bar{\theta } \sin \left(\bar{\alpha }+\beta_l\right)}{\bar{\rho } y_l}+\frac{r_l^2 \csc ^2\bar{\theta } \sin \left(2 \left(\bar{\alpha
   }+\beta_l\right)\right)}{2 \bar{\rho }^2 y_l^2}+\CO\left(\frac{1}{\bar{\rho }^3}\right),
\eeq
where $(\vec x_l)_1=r_l \cos \beta_l, (\vec x_l)_2 = - r_l \sin \beta_l$. Now we can use the first equation here and the cut-off \eqref{eq:cutoffinalpha} to derive the FG cut-off in the $\bar \rho$ coordinate:
\beq 
\bar \rho _{cut-off} &=\frac{1}{\epsilon  y_l}-\frac{r_l \sin \bar{\theta } \cos \left(\bar{\alpha }+\beta _l\right)}{y_l}+\frac{\epsilon }{16 y_l} \Big(-2 r_l^2 \left(\cos \left(2 \bar{\theta }\right)-2
   \sin ^2\bar{\theta } \cos \left(2 \left(\bar{\alpha }+\beta _l\right)\right)+3\right)+\\
   &-8 y_l^2 \cos ^2\bar{\theta }+12 \sin ^2\bar{\theta }
   \left(2 m_{211} \sin \left(2 \bar{\alpha }\right)+\left(m_{220}-m_{202}\right) \cos \left(2 \bar{\alpha }\right)\right)+\\
   &+\left(-6 m_{202}-6 m_{220}+3 m_{400}+1\right) \cos \left(2
   \bar{\theta }\right)+2 m_{202}+2 m_{220}-m_{400}+1\Big)+ \CO\left(\epsilon ^2\right),
\eeq

Then, 
\beq 
&{1 \over 2} \int_X \; y \; z_l \, dx_1 \, dx_2 \, dy = \frac{\pi  y_l^2}{4 \epsilon ^2}-\frac{1}{24} \pi  y_l^2 \left(4 \left( (\vec x_l)_1^2 + (\vec x_l)_2^2 \right)+4 y_l^2-1\right)+\CO\left(\epsilon ^1\right),
\eeq
the moments drop out in the angular integrals, but we see that after summing all the integrals they appear again: 
\beq 
\label{eq:secondtermI1}
\sum_{l=1}^M \int_X \; y \, z_l \, dx_1 \, dx_2 \, dy = \frac{\pi }{4\varepsilon^2}+\frac{\pi }{24} \, \left(1 - 4\left(m_{220}+m_{202}+m_{400}\right)\right)+\CO\left(\epsilon\right).
\eeq

Using \eqref{eq:firstermI1} and \eqref{eq:secondtermI1} we can write
\beq 
I_1 &= - { 4 \over \kappa^2 } \vol (AdS_3) \; \vol (S^3) \; \vol (S^1) \; \Big(\frac{\pi }{8 \epsilon ^4}+\frac{\pi }{32 \epsilon ^2} \left(1+2 m_{202}+2 m_{220}-m_{400} \right)+\\
&+\frac{\pi}{1536}  \Big(-288 m_{202}^2+144 m_{220} m_{202}+108 m_{400} m_{202}+12 m_{202}-720
   m_{211}^2-288 m_{220}^2+\\
   &-27 m_{400}^2+48 m_{204}+12 m_{220}+96 m_{222}+48 m_{240}+108 m_{220} m_{400}-6 m_{400}-96 m_{402}+\\
   &-96 m_{420}+16 m_{600}-7\Big)  \Big)+O\left(\epsilon ^2\right)
\eeq

The remaining integrals are boundary terms. We start with $I_2$ \eqref{eq:integrals}
\beq 
I_2&=- { 4 \over \kappa^2 } \vol (AdS_3) \; \vol (S^3) \; \vol (S^1) \; \int_X \d_I u^I dx_1\, dx_2 \, dy = \\
&=- { 4 \over \kappa^2 } \vol (AdS_3) \; \vol (S^3) \; \vol (S^1) \; \int_{\partial X} \tilde n_I u^I \; \sqrt{\gamma} \; d^2x,\\
\eeq
where $u_I$ is given by \eqref{eq:ui}, and $\gamma$ is the induced metric on $\d X$.  

Note that $\partial X$ is the two-dimensional regulated boundary\footnote{There is also a boundary at $y=0$ but the integrand vanishes there. Note that this is a boundary of $X$ but not a boundary of the spacetime, since the $S^3$ shrinks smoothly and it corresponds to points in the interior. } given by the cut-off surface $\rho = \rho_{c}(\theta,\alpha)$. In the $\{\rho,\theta,\alpha\}$ coordinates on $X$ \eqref{eq:metricX}, the vector $\tilde n$ takes the following form
\beq 
\tilde n_\rho = \frac{1}{\mathcal{D}}, \quad \tilde n_\theta = - \frac{1}{\mathcal{D}} \; \d_\theta \rho_{c}(\epsilon,\theta, \alpha) , \quad \tilde n_\alpha = - \frac{1}{\mathcal{D}} \; \d_\alpha \rho_{c}(\epsilon,\theta, \alpha), \quad 
\eeq
where $\CD$ is the norm of the vector $(1,-\d_\theta \rho_{c}, - d_\alpha \rho_c)$ with respect to the metric \eqref{eq:metricX}: 
\beq 
\CD^2 = \frac{\rho^2 + 1+(\d_\theta \rho_c)^2}{\rho^2 + \sin^2 \theta}  + \frac{(\d_\alpha \rho_c)^2}{\rho^2  \sin^2 \theta} .
\eeq

The induced metric is given by 
\beq 
\gamma_{ab} = g_{IJ} \, {\d x^I \over \d y_a} \, {\d x^J \over \d y_b},
\eeq
where $g_{IJ}$ is the metric on $X$ \eqref{eq:metricX}, $x^I$ are coordinates $\{\rho,\theta,\alpha\}$ on $X$ and $y_a$ are coordinates  $\{\theta,\alpha\}$ on $\d X$. Then, 
\beq
\sqrt{\gamma} = \sqrt{g} \; \CD^2  
\eeq
And after integrating the angular variables:
\beq 
\label{eq:bulk 2nd calc}
I_2&=- { 4 \over \kappa^2 } \vol (AdS_3) \; \vol (S^3) \; \vol (S^1) \; \int_{\partial X} \tilde n_I \; u^I \; \sqrt{\gamma} \; d^2x= \\
&=- { 4 \over \kappa^2 } \vol (AdS_3) \; \vol (S^3) \; \vol (S^1) \; \Big(- \frac{\pi}{16\varepsilon^4} +  \frac{\pi}{64\varepsilon^2}\left(1+2m_{220}+2m_{202}-m_{400}\right) +\\
&+\frac{\pi}{3072}\Big(-51  -100 m_{220}  -100 m_{202}+50m_{400} +\\
&+72 \left(m_{220}^2 +m_{202}^2\right)+144 m_{211}^2-36 \left(m_{220}+m_{202}\right)m_{400}+9 m_{400}^2 +\\
& +48 \left(m_{240}+ m_{204}\right)    + 96 \left(m_{222} -m_{402}-m_{420}\right)+16m_{600} \Big) \Big) + \CO (\epsilon^1).
\eeq

Lastly, we need to include the Gibbons-Hawking-York term, evaluated in the nine-dimensional boundary: 
\beq 
I_3=& {1 \over \kappa^2} \int d^9x \sqrt{h} \, K .
\eeq

In appendix \ref{appendixarea} we compute $\sqrt{h}$ \eqref{eq:sqrth}, and the leading $\epsilon$ divergence is $\CO (\epsilon^{-4})$, so we need the trace of the extrinsic curvature of the nine-dimensional boundary up to order $\CO (\epsilon^{4})$.
\beq 
K = \nabla_\mu n^\mu
\eeq
We omit the explicit expression of $K$ in terms of the momenta because it is too long. The integrand is given by 
\beq
\sqrt{h} \, K  &= \frac{4 \sin \theta \cos ^3\theta}{\epsilon ^4}+\frac{1}{\epsilon ^2} \Big(\frac{9}{4} \sin ^32 \theta \left(2 \sin 2 \alpha m_{211}+\cos 2 \alpha
   \left(m_{220}-m_{202}\right)\right)+\\
&+\frac{1}{2} \sin \theta \cos ^3\theta \left(-9 \cos 2 \theta \left(2 m_{202}+2
   m_{220}-m_{400}+1\right)+6 m_{202}+6 m_{220}-3 m_{400}+5\right)\Big)+\\
&+\frac{28 \sin ^2\theta \cos ^3\theta}{3 \epsilon
   } \Big(-3  \cos 2 \theta \left( \cos \alpha\left(m_{212}+m_{230}-m_{410}\right)- \sin \alpha  \left(m_{203}+m_{221}-m_{401}\right)\right)+\\
   &+2 \sin ^2\theta \left(\cos 3 \alpha \left(m_{230}-3 m_{212}\right)-\sin (3
   \alpha ) \left(m_{203}-3 m_{221}\right)\right)\Big)+\\
&+\frac{1}{64} \Bigg[256 \cos ^3\theta \sin 4 \alpha \left(9 m_{211} \left(m_{202}-m_{220}\right)+20 \left(m_{231}-m_{213}\right)\right)
   \sin ^5\theta+\\
&-64 \cos 4 \alpha \cos ^3\theta \left(9 \left(m_{202}-m_{220}\right){}^2-4 \left(9 m_{211}^2+5 m_{204}+5 \left(m_{240}-6
   m_{222}\right)\right)\right) \sin ^5\theta+\\
&+4 \sin 2 \alpha \sin ^32 \theta \Big(-32 \left(m_{213}+m_{231}\right)+9 m_{211}
   \left(m_{400}-2 \left(m_{202}+m_{220}-1\right)\right)+48 m_{411}+\\
&+4 \cos 2 \theta \left(-40 \left(m_{213}+m_{231}\right)+3 m_{211} \left(6
   m_{202}+6 m_{220}-3 m_{400}-5\right)+60 m_{411}\right)\Big)+\\
&-2 \cos 2 \alpha \sin ^32 \theta \big(-18 m_{202}^2+9
   \left(m_{400}+2\right) m_{202}-32 m_{204}+9 m_{220} \left(2 m_{220}-m_{400}-2\right)+\\
&+4 \cos 2 \theta \big(18 m_{202}^2-3 \left(3
   m_{400}+5\right) m_{202}-40 m_{204}+3 m_{220} \left(-6 m_{220}+3 m_{400}+5\right)+\\
&+20 \left(2 m_{240}+3 m_{402}-3 m_{420}\right)\big)+16
   \left(2 m_{240}+3 m_{402}-3 m_{420}\right)\big)+\\
&+\frac{1}{16} \Bigg(\sin 2 \theta \Big(-144 m_{211}^2-192 m_{204}-384 m_{222}-192
   m_{240}+36 m_{202} \left(4 m_{220}-m_{400}-1\right)+\\
&-36 m_{220} \left(m_{400}+1\right)+9 m_{400} \left(m_{400}+2\right)+384 m_{402}+384 m_{420}-64 m_{600}+37\Big)+\\
&-3 \sin (6 \theta ) \big(-144 m_{211}^2-192 m_{204}-384 m_{222}-192 m_{240}+36 m_{202} \left(4
   m_{220}-m_{400}-1\right)+\\
&-36 m_{220} \left(m_{400}+1\right)+9 m_{400} \left(m_{400}+2\right)+384 m_{402}+384 m_{420}-64 m_{600}+37\big)+\\
&-4
   \sin (8 \theta ) \big(54 m_{202}^2+12 \left(3 m_{220}-3 m_{400}-5\right) m_{202}+72 m_{211}^2+54 m_{220}^2+9 m_{400}^2+\\
&-120 m_{204}-240
   m_{222}-120 m_{240}+30 m_{400}-12 m_{220} \left(3 m_{400}+5\right)+\\
&+240 m_{402}+240 m_{420}-40 m_{600}+1\big)+\\
&+4 \sin 4 \theta \big(108
   m_{202}^2-12 \left(6 m_{220}+3 m_{400}+7\right) m_{202}+288 m_{211}^2+108 m_{220}^2+\\
&+9 m_{400}^2-48 m_{204}-96 m_{222}-48 m_{240}+42 m_{400}-12
   m_{220} \left(3 m_{400}+7\right)+\\
&+96 m_{402}+96 m_{420}-16 m_{600}-35\big)\Bigg)\Bigg]+O\left(\epsilon ^1\right)
\eeq

Surprisingly, after integrating over the angular coordinates all the dependence on the momenta vanishes:
\beq
I_3=& {1 \over \kappa^2} \int d^9x \sqrt{h} \, K = {\pi \over \kappa^2}  \vol (AdS_3) \; \vol (S^3) \; \vol (S^1) \; \left( {4 \over \epsilon^4} + {1 \over \epsilon^2} \right) +\CO (\epsilon^1)
\eeq

\section{Area of the regulated boundary}
\label{appendixarea}

We want to compute the area of the regulated boundary, which is located at 
\beq 
\rho = \rho_{cut-off}( \epsilon, \theta, \phi, \psi),
\eeq
with the cut-off function found in \eqref{eq:cutoff}. The induced metric in the boundary is
\beq 
h_{\mu\, \nu} = g_{\mu \nu} - n_\mu \, n_\nu, \quad h_{ab} = g_{\mu \, \nu} \, e^{\mu}_a \, e^{\nu}_b
\eeq
where $n_\mu$ is a unit vector normal to the boundary and $e^\mu_a = \frac{\d x^\mu}{\d y^a}$, where $\{x_\mu\}$ are the coordinates of the 10 dimensional spacetime and $\{y_a\}$ are coordinates in the 9 dimensional boundary, given by parametric equations $x^\mu = x^\mu(y^a)$.

We will describe the 10 dimensional space using the coordinates $\{\rho, Z, X , Y, \psi, \phi_1, \phi_2,\phi_3, \theta, \alpha\}$, which corresponds to starting with the metric in \eqref{eq:metricbubblingsurface} and doing the change of coordinates \eqref{changeofcoords} with $\alpha = \phi + \psi$. $\{Z,X,Y\}$ are $AdS_3$ coordinates and $\{\phi_1,\phi_2,\phi_3\}$ are coordinates on $S^3$. 


Note that the cut-off function is then written like 
\beq
\label{eq:cutoffinalpha}
\rho_{cut-off}&(\epsilon,\psi,\theta,\alpha) =  \frac{1}{\epsilon} + \frac{F_\rho^{(2)}-1}{4}\,\epsilon +  \frac{F_\rho^{(3)}}{6}\, \epsilon^2     \\
& + \frac{16\left(F_\rho^{(4)}-F_\rho^{(2)}\left(F_\rho^{(2)}-1\right)\right)-\left(\partial_\theta F_\rho^{(2)}\right)^2- \left(\partial_\alpha F_\rho^{(2)}\right)^2 \csc^2\theta}{128}\,  \epsilon^3 +O\left(\epsilon^4\right),
\eeq
and since $F_\rho^{(2,3,4)}$ depend only on $(\theta, \alpha)$, the cut-off function is also $\psi$ independent. Then, the regulated boundary is at $\rho = \rho_{c}(\theta,\alpha)$.

The only nonzero components of the unit normal vector to the boundary are 
\beq 
n_\rho = \frac{1}{\mathcal{N}}, \quad n_\theta = - \frac{1}{\mathcal{N}} \; \d_\theta \rho_{c}(\epsilon,\theta, \alpha) , \quad n_\alpha = - \frac{1}{\mathcal{N}} \; \d_\alpha \rho_{c}(\epsilon,\theta, \alpha), \quad 
\eeq
where $\CN$ is the norm of the vector $(1,-\d_\theta \rho_{c}, - d_\alpha \rho_c, \vec 0)$. 

Note that we only need the part of the metric involving the coordinates $\{\rho, \theta, \alpha, \psi\}$ (there are no off-diagonal components mixing any of these coordinates with the rest). This part of the metric looks like 
\beq 
&\frac{2 y}{\sqrt{4z^2-1}} \Bigg(\frac{4z^2 -1}{4y^2}  \frac{\rho^2 + \sin^2 \theta}{\rho^2 + 1} \, d\rho^2+  d \psi^2+\left(\frac{4z^2 -1}{4y^2}\left( \rho^2 + \sin^2 \theta \right) +  V_\theta^2 \right)d\theta^2 + \\
+&\left(\frac{4z^2 -1}{4y^2}  \rho^2 \sin^2\theta + \left(V_\alpha -\frac{1}{2}\right)^2 \right) \, d\alpha^2+\\
&+ 2 V_\theta \, d\psi \, d\theta + (2V_\alpha -1) \, d\psi \, d\alpha+  V_\theta (2V_\alpha -1) \, d\theta \, d\alpha \Bigg)
\eeq
where we have explicitly brought $V$ to a gauge such that $V_\rho=0$. With this, we can compute the normalization factor 
\beq 
\CN^2 = \frac{2 \sqrt{\rho^2 +1} \, \cos \theta}{\sqrt{4z^2-1}} \left(\frac{\rho^2 + 1}{\rho^2 + \sin^2 \theta} + \frac{(\d_\theta \rho_c)^2}{\rho^2 + \sin^2 \theta} + \frac{(\d_\alpha \rho_c)^2}{\rho^2  \sin^2 \theta} \right),
\eeq
and the determinant of the induced metric: 
\beq 
\label{eq:dethexpression}
\det h &= {1 \over 4} y^5 \left( 4z^2 -1\right)^{1/2} (\rho^2+\sin^2\theta) \rho^2 \sin^2 \theta \times \\
&\times \left( 1 + {(\d_\theta \rho_c)^2 \over \rho^2 + 1} + { \rho^2 + \sin^2\theta \over \rho^2 \sin^2 \theta  (\rho^2 + 1)} (\d_\alpha \rho_c)^2\right) {1 \over Z^6} \, \sin^4 \phi_2 \, \sin^2 \phi_3 \Bigg|_{\rho = \rho_c}.
\eeq
We could also compute $
\sqrt{h}$ from the large $\rho$ expansion of the metric \eqref{eq:largeRhoMetric}:
\beq 
\sqrt{h}&=\sin \theta \cos ^3\theta \Big(\frac{1}{\epsilon ^4}+\frac{1}{2 \epsilon ^2} (3 F_1^{(2)}+F_2^{(2)}+3 F_3^{(2)}+F_4^{(2)}+F_5^{(2)}+2 F_\rho^{(2)}+1)+\\
&+\frac{1}{6
   \epsilon } (9 F_1^{(3)}+3 F_2^{(3)}+9 F_3^{(3)}+3 F_4^{(3)}+3 F_5^{(3)}+4 F_\rho^{^{(3)}})+\\
   &+\frac{1}{8}  \Big(3 (F_1^{(2)})^2+12 F_1^{(4)}+6 F_1^{(2)}
   (F_2^{(2)}+3 F_3^{(2)}+F_4^{(2)}+F_5^{(2)}+F_\rho^{(2)}+2)+\\
   &-(F_2^{(2)})^2+4 F_2^{(2)}+4 F_2^{(4)}+6 F_2^{(2)} F_3^{(2)}+2 F_2^{(2)}
   F_4^{(2)}+2 F_2^{(2)} F_5^{(2)}+2 F_2^{(2)} F_\rho^{(2)}+\\
   &+3 (F_3^{(2)})^2+12 F_3^{(2)}+12 F_3^{(4)}+6 F_3^{(2)} F_4^{(2)}+6 F_3^{(2)}
   F_5^{(2)}+6 F_3^{(2)} F_\rho^{(2)}-(F_4^{(2)})^2+\\
   &+4 F_4^{(2)}+4 F_4^{(4)}+2 F_4^{(2)} F_5^{(2)}+2 F_4^{(2)} F_\rho^{(2)}-(F_5^{(2)})^2+4
   F_5^{(2)}+4 F_5^{(4)}+2 F_5^{(2)} F_\rho^{(2)}+\\
   &-(F_\rho^{(2)})^2+4
   F_\rho^{(2)}+4 F_\rho^{(4)}\Big)\Big)+\CO(\epsilon ^1)
\eeq

Note however that \eqref{eq:dethexpression} only depends on $F_\rho$ and its derivatives, so the metric functions are not independent when evaluated in the cut-off surface.

After regularizing using the cut-off $\epsilon$ defined in \eqref{eq:cutoff} we find the following expression depending on the momenta:
\beq 
\label{eq:sqrth}
\sqrt{h} &= \frac{1}{\epsilon ^4}\sin \theta \cos ^3\theta+\frac{1}{8 \epsilon ^2} \Big(\frac{9}{2} \sin ^32 \theta \left(2 \sin 2 \alpha m_{211}+\cos 2 \alpha \left(m_{220}-m_{202}\right)\right)+\\
&+\sin \theta \cos ^3\theta \left(-9 \cos 2 \theta \left(2 m_{202}+2 m_{220}-m_{400}+1\right)+6 m_{202}+6 m_{220}-3 m_{400}+7\right) \Big)+\\
&+\frac{7 \sin ^2\theta \cos ^3\theta}{3 \epsilon } \Big(-3 \cos \alpha \cos 2 \theta \left(m_{212}+m_{230}-m_{410}\right)+\\
&+2 \sin ^2\theta \left(\cos 3 \alpha \left(m_{230}-3 m_{212}\right)-\sin 3 \alpha \left(m_{203}-3 m_{221}\right)\right)+\\
&-3 \sin \alpha \cos 2 \theta \left(m_{203}+m_{221}-m_{401}\right)\Big)+\\
&+\frac{1}{64} \Big[64 \cos ^3\theta \sin 4 \alpha \sin ^5\theta \left(9 m_{211}
   \left(m_{202}-m_{220}\right)+20 \left(m_{231}-m_{213}\right)\right) +\\
   &-16 \cos 4 \alpha \cos ^3\theta \sin ^5\theta \left(9 \left(m_{202}-m_{220}\right){}^2-4 \left(9 m_{211}^2+5 m_{204}+5 \left(m_{240}-6 m_{222}\right)\right)\right) +\\
   &-\cos ^3\theta \sin \theta \Big(98 m_{202}^2-20 m_{220} m_{202}-52 m_{202}+216 m_{211}^2+98 m_{220}^2+11 m_{400}^2+\cos 4 \theta+\\
   &-72 m_{204}-52 m_{220}-144 m_{222}-72 m_{240}-44 \left(m_{202}+m_{220}\right) m_{400}+26 m_{400}+144 m_{402}+\\
   &+144 m_{420}+\cos 4 \theta \big(54 m_{202}^2+12 \left(3 m_{220}-3 m_{400}-5\right) m_{202}+72 m_{211}^2+54 m_{220}^2+9 m_{400}^2+\\
   &-120 m_{204}-12 m_{220} \left(3 m_{400}+5\right)+30 \left(-8 m_{222}-4 m_{240}+m_{400}+8 \left(m_{402}+m_{420}\right)\right)-40
   m_{600}\big)+\\
   &-4 \cos 2 \theta \big(30 m_{202}^2-12 \left(m_{220}+m_{400}+3\right) m_{202}+72 m_{211}^2+30 m_{220}^2-24 m_{204}-48 m_{222}+\\
   &-24 m_{240}-12 m_{220} \left(m_{400}+3\right)+3 m_{400} \left(m_{400}+6\right)+48 m_{402}+48 m_{420}-8 m_{600}-13\big)+\\
   &-24 m_{600}-13\Big) +4 \sin 2 \alpha \sin ^32 \theta \Big(-8 \left(m_{213}+m_{231}\right)+\\
   &+3 m_{211} \left(-2 m_{202}-2 m_{220}+m_{400}+3\right)+12 m_{411}+\\
   &+\cos 2 \theta \left(-40 \left(m_{213}+m_{231}\right)+3 m_{211} \left(6 m_{202}+6 m_{220}-3 m_{400}-5\right)+60 m_{411}\right)\Big)+\\
   &-2 \cos 2 \alpha \sin ^32 \theta \Big(-6 m_{202}^2+3 \left(m_{400}+3\right) m_{202}-8 m_{204}+3 m_{220} \left(2 m_{220}-m_{400}-3\right)+\\
   &+\cos 2 \theta \big(18 m_{202}^2-3
   \left(3 m_{400}+5\right) m_{202}-40 m_{204}+3 m_{220} \left(-6 m_{220}+3 m_{400}+5\right)+\\
   &+20 \left(2 m_{240}+3 m_{402}-3 m_{420}\right)\big)+4 \left(2 m_{240}+3 m_{402}-3 m_{420}\right)\Big)\Big]+O\left(\epsilon ^1\right).
\eeq

Most of the terms vanish upon integration in the angular coordinates because of symmetry. The final result is 
\beq 
a \frac{1}{2\kappa^2\, L} \int d^9x \sqrt{h} = a \frac{1}{2\kappa^2\, L} \vol (AdS_3) \, \vol (S^3) \, \vol (S^1) \,  \pi \left( \frac{\epsilon^4}{2} + \frac{\epsilon^2}{4} + \mathcal{C} \right),
\eeq
where 
\beq 
\mathcal{C} &= \frac{1}{32} \left(-10 m_{2,0,2}^2+4 \left(m_{2,2,0}+m_{4,0,0}-1\right) m_{2,0,2}-24 m_{2,1,1}^2-10 m_{2,2,0}^2-m_{4,0,0}^2+ \right. \\
&\left.-4 m_{2,2,0}+4 m_{2,2,0} m_{4,0,0}+2 m_{4,0,0}-1\right)
\eeq

\end{appendix}

\bibliographystyle{utphys.bst}
\bibliography{references.bib}

@article{Constable:2002xt,
    author = "Constable, Neil R. and Erdmenger, Johanna and Guralnik, Zachary and Kirsch, Ingo",
    title = "{Intersecting D-3 branes and holography}",
    eprint = "hep-th/0211222",
    archivePrefix = "arXiv",
    reportNumber = "HU-EP-02-32, MIT-CTP-3316",
    doi = "10.1103/PhysRevD.68.106007",
    journal = "Phys. Rev. D",
    volume = "68",
    pages = "106007",
    year = "2003"
}

@article{fefferman1985conformal,
  title={Conformal invariants},
  author={Fefferman, C. and Graham, C. R. },
  journal={" Elie Cartan et les Mathematiques d'Aujourd'hui," Asterisque, hors serie},
  pages={95--116},
  year={1985}
}

@article{Henningson:1998gx,
    author = "Henningson, M. and Skenderis, K.",
    title = "{The Holographic Weyl anomaly}",
    eprint = "hep-th/9806087",
    archivePrefix = "arXiv",
    reportNumber = "CERN-TH-98-188, KUL-TF-98-21",
    doi = "10.1088/1126-6708/1998/07/023",
    journal = "JHEP",
    volume = "07",
    pages = "023",
    year = "1998"
}

@article{Beccaria:2023hhi,
    author = "Beccaria, Matteo and Tseytlin, Arkady A.",
    title = "{Comments on ABJM free energy on S3 at large N and perturbative expansions in M-theory and string theory}",
    eprint = "2306.02862",
    archivePrefix = "arXiv",
    primaryClass = "hep-th",
    reportNumber = "Imperial-TP-AT-2023-03",
    doi = "10.1016/j.nuclphysb.2023.116286",
    journal = "Nucl. Phys. B",
    volume = "994",
    pages = "116286",
    year = "2023"
}

@article{Gentle:2015ruo,
    author = "Gentle, Simon A. and Gutperle, Michael and Marasinou, Chrysostomos",
    title = "{Holographic entanglement entropy of surface defects}",
    eprint = "1512.04953",
    archivePrefix = "arXiv",
    primaryClass = "hep-th",
    doi = "10.1007/JHEP04(2016)067",
    journal = "JHEP",
    volume = "04",
    pages = "067",
    year = "2016"
}

@article{Kurlyand:2022vzv,
    author = "Kurlyand, S. A. and Tseytlin, A. A.",
    title = "{Type IIB supergravity action on M5\texttimes{}X5 solutions}",
    eprint = "2206.14522",
    archivePrefix = "arXiv",
    primaryClass = "hep-th",
    reportNumber = "Imperial-TP-AT-2022-03",
    doi = "10.1103/PhysRevD.106.086017",
    journal = "Phys. Rev. D",
    volume = "106",
    number = "8",
    pages = "086017",
    year = "2022"
}

@article{Giddings:2001yu,
    author = "Giddings, Steven B. and Kachru, Shamit and Polchinski, Joseph",
    title = "{Hierarchies from fluxes in string compactifications}",
    eprint = "hep-th/0105097",
    archivePrefix = "arXiv",
    reportNumber = "SLAC-PUB-8807, NSF-ITP-01-37, SU-ITP-01-16",
    doi = "10.1103/PhysRevD.66.106006",
    journal = "Phys. Rev. D",
    volume = "66",
    pages = "106006",
    year = "2002"
}

@article{DeWolfe:2002nn,
    author = "DeWolfe, Oliver and Giddings, Steven B.",
    title = "{Scales and hierarchies in warped compactifications and brane worlds}",
    eprint = "hep-th/0208123",
    archivePrefix = "arXiv",
    reportNumber = "NSF-ITP-02-71, SU-ITP-02-27",
    doi = "10.1103/PhysRevD.67.066008",
    journal = "Phys. Rev. D",
    volume = "67",
    pages = "066008",
    year = "2003"
}

@article{Gomis:2007fi,
    author = "Gomis, Jaume and Matsuura, Shunji",
    title = "{Bubbling surface operators and S-duality}",
    eprint = "0704.1657",
    archivePrefix = "arXiv",
    primaryClass = "hep-th",
    doi = "10.1088/1126-6708/2007/06/025",
    journal = "JHEP",
    volume = "06",
    pages = "025",
    year = "2007"
}

@article{Lin:2004nb,
    author = "Lin, Hai and Lunin, Oleg and Maldacena, Juan Martin",
    title = "{Bubbling AdS space and 1/2 BPS geometries}",
    eprint = "hep-th/0409174",
    archivePrefix = "arXiv",
    reportNumber = "PUPT-2136",
    doi = "10.1088/1126-6708/2004/10/025",
    journal = "JHEP",
    volume = "10",
    pages = "025",
    year = "2004"
}

@article{Henningson:1998ey,
    author = "Henningson, Mans and Skenderis, Kostas",
    editor = "Lust, D. and Otto, H. J.",
    title = "{Holography and the Weyl anomaly}",
    eprint = "hep-th/9812032",
    archivePrefix = "arXiv",
    reportNumber = "GOTEBORG-ITP-98-14, SPIN-1998-08, SPIN-1998-8",
    doi = "10.1002/(SICI)1521-3978(20001)48:1/3<125::AID-PROP125>3.0.CO;2-B",
    journal = "Fortsch. Phys.",
    volume = "48",
    pages = "125--128",
    year = "2000"
}

@article{Frey:2019fqz,
    author = "Frey, Andrew R",
    title = "{Dirac branes for Dirichlet branes: Supergravity actions}",
    eprint = "1907.12755",
    archivePrefix = "arXiv",
    primaryClass = "hep-th",
    doi = "10.1103/PhysRevD.102.046017",
    journal = "Phys. Rev. D",
    volume = "102",
    number = "4",
    pages = "046017",
    year = "2020"
}

@article{Drukker:2008wr,
    author = "Drukker, Nadav and Gomis, Jaume and Matsuura, Shunji",
    title = "{Probing N=4 SYM With Surface Operators}",
    eprint = "0805.4199",
    archivePrefix = "arXiv",
    primaryClass = "hep-th",
    reportNumber = "HU-EP-08-17",
    doi = "10.1088/1126-6708/2008/10/048",
    journal = "JHEP",
    volume = "10",
    pages = "048",
    year = "2008"
}

@article{Emparan:1999pm,
    author = "Emparan, Roberto and Johnson, Clifford V. and Myers, Robert C.",
    title = "{Surface terms as counterterms in the AdS / CFT correspondence}",
    eprint = "hep-th/9903238",
    archivePrefix = "arXiv",
    reportNumber = "DTP-99-21, UK-99-04, MCGILL-99-12, EHU-FT-9906",
    doi = "10.1103/PhysRevD.60.104001",
    journal = "Phys. Rev. D",
    volume = "60",
    pages = "104001",
    year = "1999"
}

@article{Witten:1998zw,
    author = "Witten, Edward",
    editor = "Bergstrom, L. and Lindstrom, U.",
    title = "{Anti-de Sitter space, thermal phase transition, and confinement in gauge theories}",
    eprint = "hep-th/9803131",
    archivePrefix = "arXiv",
    reportNumber = "IASSNS-HEP-98-21",
    doi = "10.4310/ATMP.1998.v2.n3.a3",
    journal = "Adv. Theor. Math. Phys.",
    volume = "2",
    pages = "505--532",
    year = "1998"
}

@article{Taylor:2001fe,
    author = "Taylor, Marika",
    title = "{Higher dimensional formulation of counterterms}",
    eprint = "hep-th/0110142",
    archivePrefix = "arXiv",
    reportNumber = "SPIN-2001-21",
    month = "10",
    year = "2001"
}

@article{Wang:2020xkc,
    author = "Wang, Yifan",
    title = "{Surface defect, anomalies and b-extremization}",
    eprint = "2012.06574",
    archivePrefix = "arXiv",
    primaryClass = "hep-th",
    doi = "10.1007/JHEP11(2021)122",
    journal = "JHEP",
    volume = "11",
    pages = "122",
    year = "2021"
}

@article{Skenderis:2007yb,
    author = "Skenderis, Kostas and Taylor, Marika",
    title = "{Anatomy of bubbling solutions}",
    eprint = "0706.0216",
    archivePrefix = "arXiv",
    primaryClass = "hep-th",
    reportNumber = "ITFA-2007-17",
    doi = "10.1088/1126-6708/2007/09/019",
    journal = "JHEP",
    volume = "09",
    pages = "019",
    year = "2007"
}

@article{Jensen:2013lxa,
    author = "Jensen, Kristan and O'Bannon, Andy",
    title = "{Holography, Entanglement Entropy, and Conformal Field Theories with Boundaries or Defects}",
    eprint = "1309.4523",
    archivePrefix = "arXiv",
    primaryClass = "hep-th",
    reportNumber = "YITP-SB-13-27, DAMTP-2013-54",
    doi = "10.1103/PhysRevD.88.106006",
    journal = "Phys. Rev. D",
    volume = "88",
    number = "10",
    pages = "106006",
    year = "2013"
}

@article{Gukov:2006jk,
    author = "Gukov, Sergei and Witten, Edward",
    title = "{Gauge Theory, Ramification, And The Geometric Langlands Program}",
    eprint = "hep-th/0612073",
    archivePrefix = "arXiv",
    month = "12",
    year = "2006"
}

@article{Gutperle:2020gez,
    author = "Gutperle, Michael and Uhlemann, Christoph F.",
    title = "{Janus on the Brane}",
    eprint = "2003.12080",
    archivePrefix = "arXiv",
    primaryClass = "hep-th",
    doi = "10.1007/JHEP07(2020)243",
    journal = "JHEP",
    volume = "07",
    pages = "243",
    year = "2020"
}

@article{Choi:2024ktc,
    author = "Choi, Changha and Gomis, Jaume and Izquierdo Garc{\'\i}a, Raquel",
    title = "{Surface operators and exact holography}",
    eprint = "2406.08541",
    archivePrefix = "arXiv",
    primaryClass = "hep-th",
    doi = "10.1007/JHEP12(2024)195",
    journal = "JHEP",
    volume = "12",
    pages = "195",
    year = "2024"
}

@article{Jensen:2018rxu,
    author = "Jensen, Kristan and O'Bannon, Andy and Robinson, Brandon and Rodgers, Ronnie",
    title = "{From the Weyl Anomaly to Entropy of Two-Dimensional Boundaries and Defects}",
    eprint = "1812.08745",
    archivePrefix = "arXiv",
    primaryClass = "hep-th",
    doi = "10.1103/PhysRevLett.122.241602",
    journal = "Phys. Rev. Lett.",
    volume = "122",
    number = "24",
    pages = "241602",
    year = "2019"
}

@article{Hartnoll:2008kx,
    author = "Hartnoll, Sean A. and Herzog, Christopher P. and Horowitz, Gary T.",
    title = "{Holographic Superconductors}",
    eprint = "0810.1563",
    archivePrefix = "arXiv",
    primaryClass = "hep-th",
    doi = "10.1088/1126-6708/2008/12/015",
    journal = "JHEP",
    volume = "12",
    pages = "015",
    year = "2008"
}

@article{Takayanagi:2011zk,
    author = "Takayanagi, Tadashi",
    title = "{Holographic Dual of BCFT}",
    eprint = "1105.5165",
    archivePrefix = "arXiv",
    primaryClass = "hep-th",
    reportNumber = "IPMU11-0091",
    doi = "10.1103/PhysRevLett.107.101602",
    journal = "Phys. Rev. Lett.",
    volume = "107",
    pages = "101602",
    year = "2011"
}

@article{Lee:2008xf,
    author = "Lee, Sung-Sik",
    title = "{A Non-Fermi Liquid from a Charged Black Hole: A Critical Fermi Ball}",
    eprint = "0809.3402",
    archivePrefix = "arXiv",
    primaryClass = "hep-th",
    doi = "10.1103/PhysRevD.79.086006",
    journal = "Phys. Rev. D",
    volume = "79",
    pages = "086006",
    year = "2009"
}

@article{Kovtun:2004de,
    author = "Kovtun, P. and Son, Dan T. and Starinets, Andrei O.",
    title = "{Viscosity in strongly interacting quantum field theories from black hole physics}",
    eprint = "hep-th/0405231",
    archivePrefix = "arXiv",
    reportNumber = "INT-PUB-04-09, UW-PT-04-04",
    doi = "10.1103/PhysRevLett.94.111601",
    journal = "Phys. Rev. Lett.",
    volume = "94",
    pages = "111601",
    year = "2005"
}

@inbook{Gukov:2014gja,
    author = "Gukov, Sergei",
    title = "{Surface operators.}",
    booktitle = "{New Dualities of Supersymmetric Gauge Theories}",
    eprint = "1412.7127",
    archivePrefix = "arXiv",
    primaryClass = "hep-th",
    doi = "10.1007/978-3-319-18769-3_8",
    pages = "223--259",
    year = "2016",
    publisher = "Springer"
}

@article{Gukov:2008sn,
    author = "Gukov, Sergei and Witten, Edward",
    title = "{Rigid Surface Operators}",
    eprint = "0804.1561",
    archivePrefix = "arXiv",
    primaryClass = "hep-th",
    doi = "10.4310/ATMP.2010.v14.n1.a3",
    journal = "Adv. Theor. Math. Phys.",
    volume = "14",
    number = "1",
    pages = "87--178",
    year = "2010"
}

@article{Graham:1999pm,
    author = "Graham, C. Robin and Witten, Edward",
    title = "{Conformal anomaly of submanifold observables in AdS / CFT correspondence}",
    eprint = "hep-th/9901021",
    archivePrefix = "arXiv",
    doi = "10.1016/S0550-3213(99)00055-3",
    journal = "Nucl. Phys. B",
    volume = "546",
    pages = "52--64",
    year = "1999"
}

@article{Henningson:1999xi,
    author = "Henningson, Mans and Skenderis, Kostas",
    title = "{Weyl anomaly for Wilson surfaces}",
    eprint = "hep-th/9905163",
    archivePrefix = "arXiv",
    reportNumber = "SPIN-1999-12, GOTEBORG-ITP-99-05",
    doi = "10.1088/1126-6708/1999/06/012",
    journal = "JHEP",
    volume = "06",
    pages = "012",
    year = "1999"
}

@article{Schwimmer:2008yh,
    author = "Schwimmer, A. and Theisen, S.",
    title = "{Entanglement Entropy, Trace Anomalies and Holography}",
    eprint = "0802.1017",
    archivePrefix = "arXiv",
    primaryClass = "hep-th",
    doi = "10.1016/j.nuclphysb.2008.04.015",
    journal = "Nucl. Phys. B",
    volume = "801",
    pages = "1--24",
    year = "2008"
}

@article{Lin:2005nh,
    author = "Lin, Hai and Maldacena, Juan Martin",
    title = "{Fivebranes from gauge theory}",
    eprint = "hep-th/0509235",
    archivePrefix = "arXiv",
    reportNumber = "PUPT-2172",
    doi = "10.1103/PhysRevD.74.084014",
    journal = "Phys. Rev. D",
    volume = "74",
    pages = "084014",
    year = "2006"
}

@article{Papadimitriou:2004rz,
    author = "Papadimitriou, Ioannis and Skenderis, Kostas",
    title = "{Correlation functions in holographic RG flows}",
    eprint = "hep-th/0407071",
    archivePrefix = "arXiv",
    reportNumber = "ITFA-2004-23",
    doi = "10.1088/1126-6708/2004/10/075",
    journal = "JHEP",
    volume = "10",
    pages = "075",
    year = "2004"
}

@article{Estes:2012nx,
    author = "Estes, John and O'Bannon, Andy and Tsatis, Efstratios and Wrase, Timm",
    title = "{Holographic Wilson Loops, Dielectric Interfaces, and Topological Insulators}",
    eprint = "1210.0534",
    archivePrefix = "arXiv",
    primaryClass = "hep-th",
    reportNumber = "CCTP-2012-18, DAMTP-2012-63, IMPERIAL-TP-2012-JE-01, SU-ITP-12-25",
    doi = "10.1103/PhysRevD.87.106005",
    journal = "Phys. Rev. D",
    volume = "87",
    number = "10",
    pages = "106005",
    year = "2013"
}

@article{Gentle:2014lva,
    author = "Gentle, Simon A. and Gutperle, Michael",
    title = "{Entanglement entropy of Wilson loops: Holography and matrix models}",
    eprint = "1407.5629",
    archivePrefix = "arXiv",
    primaryClass = "hep-th",
    doi = "10.1103/PhysRevD.90.066011",
    journal = "Phys. Rev. D",
    volume = "90",
    number = "6",
    pages = "066011",
    year = "2014"
}

@article{Maldacena:1997re,
    author = "Maldacena, Juan Martin",
    title = "{The Large $N$ limit of superconformal field theories and supergravity}",
    eprint = "hep-th/9711200",
    archivePrefix = "arXiv",
    reportNumber = "HUTP-97-A097, HUTP-98-A097",
    doi = "10.4310/ATMP.1998.v2.n2.a1",
    journal = "Adv. Theor. Math. Phys.",
    volume = "2",
    pages = "231--252",
    year = "1998"
}

@article{Gubser:1998bc,
    author = "Gubser, S. S. and Klebanov, Igor R. and Polyakov, Alexander M.",
    title = "{Gauge theory correlators from noncritical string theory}",
    eprint = "hep-th/9802109",
    archivePrefix = "arXiv",
    reportNumber = "PUPT-1767",
    doi = "10.1016/S0370-2693(98)00377-3",
    journal = "Phys. Lett. B",
    volume = "428",
    pages = "105--114",
    year = "1998"
}

@article{Witten:1998qj,
    author = "Witten, Edward",
    title = "{Anti de Sitter space and holography}",
    eprint = "hep-th/9802150",
    archivePrefix = "arXiv",
    reportNumber = "IASSNS-HEP-98-15",
    doi = "10.4310/ATMP.1998.v2.n2.a2",
    journal = "Adv. Theor. Math. Phys.",
    volume = "2",
    pages = "253--291",
    year = "1998"
}

@article{Kraus:1999di,
    author = "Kraus, Per and Larsen, Finn and Siebelink, Ruud",
    title = "{The gravitational action in asymptotically AdS and flat space-times}",
    eprint = "hep-th/9906127",
    archivePrefix = "arXiv",
    reportNumber = "EFI-99-29, KUL-TF-99-23",
    doi = "10.1016/S0550-3213(99)00549-0",
    journal = "Nucl. Phys. B",
    volume = "563",
    pages = "259--278",
    year = "1999"
}

@article{Papadimitriou:2004ap,
    author = "Papadimitriou, Ioannis and Skenderis, Kostas",
    editor = "Biquard, O.",
    title = "{AdS / CFT correspondence and geometry}",
    eprint = "hep-th/0404176",
    archivePrefix = "arXiv",
    reportNumber = "ITFA-2004-17",
    doi = "10.4171/013-1/4",
    journal = "IRMA Lect. Math. Theor. Phys.",
    volume = "8",
    pages = "73--101",
    year = "2005"
}

@article{Skenderis:2002wp,
    author = "Skenderis, Kostas",
    editor = "de Wit, B. and Vandoren, S.",
    title = "{Lecture notes on holographic renormalization}",
    eprint = "hep-th/0209067",
    archivePrefix = "arXiv",
    reportNumber = "PUTP-2047",
    doi = "10.1088/0264-9381/19/22/306",
    journal = "Class. Quant. Grav.",
    volume = "19",
    pages = "5849--5876",
    year = "2002"
}

@article{Skenderis:2006uy,
    author = "Skenderis, Kostas and Taylor, Marika",
    title = "{Kaluza-Klein holography}",
    eprint = "hep-th/0603016",
    archivePrefix = "arXiv",
    reportNumber = "ITFA-2006-04",
    doi = "10.1088/1126-6708/2006/05/057",
    journal = "JHEP",
    volume = "05",
    pages = "057",
    year = "2006"
}

@article{Taylor:2001pp,
    author = "Taylor, Marika",
    title = "{Anomalies, counterterms and the N=0 Polchinski-Strassler solutions}",
    eprint = "hep-th/0103162",
    archivePrefix = "arXiv",
    month = "3",
    year = "2001"
}

@article{Duff:1984hn,
    author = "Duff, M. J. and Nilsson, B. E. W. and Pope, C. N. and Warner, N. P.",
    title = "{On the Consistency of the {Kaluza-Klein} Ansatz}",
    reportNumber = "Imperial/TP/83-84/56",
    doi = "10.1016/0370-2693(84)91558-2",
    journal = "Phys. Lett. B",
    volume = "149",
    pages = "90--94",
    year = "1984"
}

@article{Turton:2024afd,
    author = "Turton, David and Tyukov, Alexander",
    title = "{Four-point correlators in $ \mathcal{N} $ = 4 SYM from AdS$_{5}$ bubbling geometries}",
    eprint = "2408.16834",
    archivePrefix = "arXiv",
    primaryClass = "hep-th",
    doi = "10.1007/JHEP10(2024)244",
    journal = "JHEP",
    volume = "10",
    pages = "244",
    year = "2024"
}

@article{Aprile:2024lwy,
    author = "Aprile, Francesco and Giusto, Stefano and Russo, Rodolfo",
    title = "{Holographic correlators with BPS bound states in $\mathcal{N} = 4$ SYM}",
    eprint = "2409.12911",
    archivePrefix = "arXiv",
    primaryClass = "hep-th",
    doi = "10.1103/PhysRevLett.134.091602",
    journal = "Phys. Rev. Lett.",
    volume = "134",
    number = "9",
    pages = "091602",
    year = "2025"
}

@article{Gomis:2008qa,
    author = "Gomis, Jaume and Matsuura, Shunji and Okuda, Takuya and Trancanelli, Diego",
    title = "{Wilson loop correlators at strong coupling: From matrices to bubbling geometries}",
    eprint = "0807.3330",
    archivePrefix = "arXiv",
    primaryClass = "hep-th",
    reportNumber = "NSF-KITP-08-96",
    doi = "10.1088/1126-6708/2008/08/068",
    journal = "JHEP",
    volume = "08",
    pages = "068",
    year = "2008"
}

@article{Gomis:2006cu,
    author = "Gomis, Jaume and Romelsberger, Christian",
    title = "{Bubbling Defect CFT's}",
    eprint = "hep-th/0604155",
    archivePrefix = "arXiv",
    doi = "10.1088/1126-6708/2006/08/050",
    journal = "JHEP",
    volume = "08",
    pages = "050",
    year = "2006"
}

@article{Yamaguchi:2006te,
    author = "Yamaguchi, Satoshi",
    title = "{Bubbling geometries for half BPS Wilson lines}",
    eprint = "hep-th/0601089",
    archivePrefix = "arXiv",
    doi = "10.1142/S0217751X07035070",
    journal = "Int. J. Mod. Phys. A",
    volume = "22",
    pages = "1353--1374",
    year = "2007"
}

@article{Lunin:2006xr,
    author = "Lunin, Oleg",
    title = "{On gravitational description of Wilson lines}",
    eprint = "hep-th/0604133",
    archivePrefix = "arXiv",
    doi = "10.1088/1126-6708/2006/06/026",
    journal = "JHEP",
    volume = "06",
    pages = "026",
    year = "2006"
}

@article{DHoker:2007mci,
    author = "D'Hoker, Eric and Estes, John and Gutperle, Michael",
    title = "{Gravity duals of half-BPS Wilson loops}",
    eprint = "0705.1004",
    archivePrefix = "arXiv",
    primaryClass = "hep-th",
    reportNumber = "UCLA-07-TEP-11",
    doi = "10.1088/1126-6708/2007/06/063",
    journal = "JHEP",
    volume = "06",
    pages = "063",
    year = "2007"
}

@article{DHoker:2007zhm,
    author = "D'Hoker, Eric and Estes, John and Gutperle, Michael",
    title = "{Exact half-BPS Type IIB interface solutions. I. Local solution and supersymmetric Janus}",
    eprint = "0705.0022",
    archivePrefix = "arXiv",
    primaryClass = "hep-th",
    reportNumber = "UCLA-07-TEP-09",
    doi = "10.1088/1126-6708/2007/06/021",
    journal = "JHEP",
    volume = "06",
    pages = "021",
    year = "2007"
}

@article{DHoker:2007hhe,
    author = "D'Hoker, Eric and Estes, John and Gutperle, Michael",
    title = "{Exact half-BPS Type IIB interface solutions. II. Flux solutions and multi-Janus}",
    eprint = "0705.0024",
    archivePrefix = "arXiv",
    primaryClass = "hep-th",
    reportNumber = "UCLA-07-TEP-10",
    doi = "10.1088/1126-6708/2007/06/022",
    journal = "JHEP",
    volume = "06",
    pages = "022",
    year = "2007"
}

@article{DHoker:2008rje,
    author = "D'Hoker, Eric and Estes, John and Gutperle, Michael and Krym, Darya",
    title = "{Exact Half-BPS Flux Solutions in M-theory II: Global solutions asymptotic to AdS(7) x S**4}",
    eprint = "0810.4647",
    archivePrefix = "arXiv",
    primaryClass = "hep-th",
    reportNumber = "UCLA-08-TEP-29, CPHT-RR079-1008",
    doi = "10.1088/1126-6708/2008/12/044",
    journal = "JHEP",
    volume = "12",
    pages = "044",
    year = "2008"
}

@article{DHoker:2008lup,
    author = "D'Hoker, Eric and Estes, John and Gutperle, Michael and Krym, Darya",
    title = "{Exact Half-BPS Flux Solutions in M-theory. I: Local Solutions}",
    eprint = "0806.0605",
    archivePrefix = "arXiv",
    primaryClass = "hep-th",
    reportNumber = "UCLA-08-TEP-16",
    doi = "10.1088/1126-6708/2008/08/028",
    journal = "JHEP",
    volume = "08",
    pages = "028",
    year = "2008"
}

@article{Jensen:2015swa,
    author = "Jensen, Kristan and O'Bannon, Andy",
    title = "{Constraint on Defect and Boundary Renormalization Group Flows}",
    eprint = "1509.02160",
    archivePrefix = "arXiv",
    primaryClass = "hep-th",
    reportNumber = "OUTP-15-19P, YITP-SB-15-33",
    doi = "10.1103/PhysRevLett.116.091601",
    journal = "Phys. Rev. Lett.",
    volume = "116",
    number = "9",
    pages = "091601",
    year = "2016"
}

@article{Bianchi:2015liz,
    author = "Bianchi, Lorenzo and Meineri, Marco and Myers, Robert C. and Smolkin, Michael",
    title = "{R{\'e}nyi entropy and conformal defects}",
    eprint = "1511.06713",
    archivePrefix = "arXiv",
    primaryClass = "hep-th",
    reportNumber = "DESY-15-229",
    doi = "10.1007/JHEP07(2016)076",
    journal = "JHEP",
    volume = "07",
    pages = "076",
    year = "2016"
}

@article{Chalabi:2020iie,
    author = "Chalabi, Adam and O'Bannon, Andy and Robinson, Brandon and Sisti, Jacopo",
    title = "{Central charges of 2d superconformal defects}",
    eprint = "2003.02857",
    archivePrefix = "arXiv",
    primaryClass = "hep-th",
    doi = "10.1007/JHEP05(2020)095",
    journal = "JHEP",
    volume = "05",
    pages = "095",
    year = "2020"
}

@article{Klebanov:1998hh,
    author = "Klebanov, Igor R. and Witten, Edward",
    title = "{Superconformal field theory on three-branes at a Calabi-Yau singularity}",
    eprint = "hep-th/9807080",
    archivePrefix = "arXiv",
    reportNumber = "IASSNS-HEP-98-64, PUPT-1804",
    doi = "10.1016/S0550-3213(98)00654-3",
    journal = "Nucl. Phys. B",
    volume = "536",
    pages = "199--218",
    year = "1998"
}

@article{Gauntlett:2004zh,
    author = "Gauntlett, Jerome P. and Martelli, Dario and Sparks, James and Waldram, Daniel",
    title = "{Supersymmetric AdS(5) solutions of M theory}",
    eprint = "hep-th/0402153",
    archivePrefix = "arXiv",
    reportNumber = "IMPERIAL-TP-03-04-6",
    doi = "10.1088/0264-9381/21/18/005",
    journal = "Class. Quant. Grav.",
    volume = "21",
    pages = "4335--4366",
    year = "2004"
}

@article{Gauntlett:2004yd,
    author = "Gauntlett, Jerome P. and Martelli, Dario and Sparks, James and Waldram, Daniel",
    title = "{Sasaki-Einstein metrics on S**2 x S**3}",
    eprint = "hep-th/0403002",
    archivePrefix = "arXiv",
    reportNumber = "IMPERIAL-TP-3-04-8",
    doi = "10.4310/ATMP.2004.v8.n4.a3",
    journal = "Adv. Theor. Math. Phys.",
    volume = "8",
    number = "4",
    pages = "711--734",
    year = "2004"
}

@article{Seiberg:1999xz,
    author = "Seiberg, Nathan and Witten, Edward",
    title = "{The D1 / D5 system and singular CFT}",
    eprint = "hep-th/9903224",
    archivePrefix = "arXiv",
    reportNumber = "IASSNS-HEP-99-27",
    doi = "10.1088/1126-6708/1999/04/017",
    journal = "JHEP",
    volume = "04",
    pages = "017",
    year = "1999"
}

@article{Kraus:1998hv,
    author = "Kraus, Per and Larsen, Finn and Trivedi, Sandip P.",
    title = "{The Coulomb branch of gauge theory from rotating branes}",
    eprint = "hep-th/9811120",
    archivePrefix = "arXiv",
    reportNumber = "EFI-98-56, FERMILAB-PUB-98-358-T",
    doi = "10.1088/1126-6708/1999/03/003",
    journal = "JHEP",
    volume = "03",
    pages = "003",
    year = "1999"
}

@article{Gaiotto:2009gz,
    author = "Gaiotto, Davide and Maldacena, Juan",
    title = "{The Gravity duals of N=2 superconformal field theories}",
    eprint = "0904.4466",
    archivePrefix = "arXiv",
    primaryClass = "hep-th",
    doi = "10.1007/JHEP10(2012)189",
    journal = "JHEP",
    volume = "10",
    pages = "189",
    year = "2012"
}

@article{Ryu:2006bv,
    author = "Ryu, Shinsei and Takayanagi, Tadashi",
    title = "{Holographic derivation of entanglement entropy from AdS/CFT}",
    eprint = "hep-th/0603001",
    archivePrefix = "arXiv",
    reportNumber = "NSF-KITP-06-11, NSF-KITP-06-11",
    doi = "10.1103/PhysRevLett.96.181602",
    journal = "Phys. Rev. Lett.",
    volume = "96",
    pages = "181602",
    year = "2006"
}

@article{BenettiGenolini:2023kxp,
    author = "Benetti Genolini, Pietro and Gauntlett, Jerome P. and Sparks, James",
    title = "{Equivariant Localization in Supergravity}",
    eprint = "2306.03868",
    archivePrefix = "arXiv",
    primaryClass = "hep-th",
    doi = "10.1103/PhysRevLett.131.121602",
    journal = "Phys. Rev. Lett.",
    volume = "131",
    number = "12",
    pages = "121602",
    year = "2023"
}

@article{BenettiGenolini:2023ndb,
    author = "Benetti Genolini, Pietro and Gauntlett, Jerome P. and Sparks, James",
    title = "{Equivariant localization for AdS/CFT}",
    eprint = "2308.11701",
    archivePrefix = "arXiv",
    primaryClass = "hep-th",
    doi = "10.1007/JHEP02(2024)015",
    journal = "JHEP",
    volume = "02",
    pages = "015",
    year = "2024"
}

@article{BenettiGenolini:2023yfe,
    author = "Benetti Genolini, Pietro and Gauntlett, Jerome P. and Sparks, James",
    title = "{Localizing wrapped M5-branes and gravitational blocks}",
    eprint = "2308.10933",
    archivePrefix = "arXiv",
    primaryClass = "hep-th",
    doi = "10.1103/PhysRevD.108.L101903",
    journal = "Phys. Rev. D",
    volume = "108",
    number = "10",
    pages = "L101903",
    year = "2023"
}

@article{Jiang:2024wzs,
    author = "Jiang, Hongliang and Tseytlin, Arkady A.",
    title = "{On co-dimension 2 defect anomalies in $ \mathcal{N} $ = 4 SYM and (2,0) theory via brane probes in AdS/CFT}",
    eprint = "2402.07881",
    archivePrefix = "arXiv",
    primaryClass = "hep-th",
    doi = "10.1007/JHEP07(2024)280",
    journal = "JHEP",
    volume = "07",
    pages = "280",
    year = "2024"
}

@article{BenettiGenolini:2024xeo,
    author = {Benetti Genolini, Pietro and Gauntlett, Jerome P. and Jiao, Yusheng and L{\"u}scher, Alice and Sparks, James},
    title = "{Localization of the Free Energy in Supergravity}",
    eprint = "2407.02554",
    archivePrefix = "arXiv",
    primaryClass = "hep-th",
    doi = "10.1103/PhysRevLett.133.141601",
    journal = "Phys. Rev. Lett.",
    volume = "133",
    number = "14",
    pages = "141601",
    year = "2024"
}

@article{BenettiGenolini:2024lbj,
    author = {Benetti Genolini, Pietro and Gauntlett, Jerome P. and Jiao, Yusheng and L{\"u}scher, Alice and Sparks, James},
    title = "{Equivariant localization for D = 4 gauged supergravity}",
    eprint = "2412.07828",
    archivePrefix = "arXiv",
    primaryClass = "hep-th",
    doi = "10.1007/JHEP08(2025)211",
    journal = "JHEP",
    volume = "08",
    pages = "211",
    year = "2025"
}

@misc{drukker_seminar_2025,
  author       = {Nadav Drukker},
  title        = {Rewarding defects},
  howpublished = {\url{https://www.newton.ac.uk/seminar/49161/}},
  year         = {2025},
  month        = {nov},
  day          = {11},
  event        = {Quantum field theory with boundaries, impurities, and defects}
}

@article{Bachas:2024nvh,
    author = "Bachas, Constantin and Chen, Zhongwu",
    title = "{Invariant tensions from holography}",
    eprint = "2404.14998",
    archivePrefix = "arXiv",
    primaryClass = "hep-th",
    doi = "10.1007/JHEP08(2024)028",
    journal = "JHEP",
    volume = "08",
    pages = "028",
    year = "2024",
    note = "[Erratum: JHEP 11, 022 (2024)]"
}

@article{Deser:1993yx,
    author = "Deser, Stanley and Schwimmer, A.",
    title = "{Geometric classification of conformal anomalies in arbitrary dimensions}",
    eprint = "hep-th/9302047",
    archivePrefix = "arXiv",
    reportNumber = "BRX-343, SISSA-14-93-EP",
    doi = "10.1016/0370-2693(93)90934-A",
    journal = "Phys. Lett. B",
    volume = "309",
    pages = "279--284",
    year = "1993"
}

@article{Gutperle:2017tjo,
    author = "Gutperle, Michael and Marasinou, Chrysostomos and Trivella, Andrea and Uhlemann, Christoph F.",
    title = "{Entanglement entropy vs. free energy in IIB supergravity duals for 5d SCFTs}",
    eprint = "1705.01561",
    archivePrefix = "arXiv",
    primaryClass = "hep-th",
    doi = "10.1007/JHEP09(2017)125",
    journal = "JHEP",
    volume = "09",
    pages = "125",
    year = "2017"
}

\end{document}